\documentclass[journal,draftcls,onecolumn,12pt,twoside]{IEEEtranTCOM}
\normalsize
\ifCLASSINFOpdf
\else
\fi

\usepackage{amsthm}
\usepackage{amsmath}
\usepackage{amssymb}
\usepackage{graphicx}
\usepackage{esint}
\usepackage{amsfonts}
\usepackage{cite}
\usepackage{balance}
\usepackage{caption}
\usepackage{subcaption}
\usepackage{epstopdf}

\makeatletter

\theoremstyle{plain}
\newtheorem{thm}{\protect\theoremname}

\makeatother

\providecommand{\theoremname}{Theorem}

\DeclareMathOperator{\Q}{Q}
\DeclareMathOperator{\IBO}{IBO}
\DeclareMathOperator{\card}{card}
\DeclareMathOperator{\sign}{sign}

\begin{document}
%\markboth{Submitted to \textit{IEEE Wireless Communications Letters}}{Karas, Karagiannidis, Smart Decode-and-Forward Relaying with Polar Codes}
\title{Spectrum Sensing Under Hardware Constraints}

\author{Alexandros~--~Apostolos~A.~Boulogeorgos,~\IEEEmembership{Student~Member,~IEEE,}~Nestor D. Chatzidiamantis, \IEEEmembership{Member, IEEE,} and George K. Karagiannidis, \IEEEmembership{Fellow, IEEE} %and Leonidas Georgiadis, \IEEEmembership{Senior, IEEE,}
\thanks{This work was presented in part at IEEE ICC 2015 - Seventh Workshop on Cooperative and Cognitive Networks (CoCoNet7). }
\thanks{A.-A. A. Boulogeorgos, N. D. Chatzidiamantis and G. K. Karagiannidis are with the Department of Electrical and Computer Engineering, Aristotle University of Thessaloniki, 54 124, Thessaloniki, Greece (e-mail:{ \{ampoulog, nestoras, geokarag\}@auth.gr}).
}

}
\maketitle	

\begin{abstract}

Direct-conversion radio (DCR) receivers can offer highly integrated low-cost hardware solutions for spectrum sensing in cognitive radio systems. However, DCR receivers are susceptible to radio frequency (RF) impairments, such as in-phase and quadrature-phase imbalance, low-noise amplifier nonlinearities and phase noise, which limit the spectrum sensing capabilities.
In this paper, we investigate the joint effects of RF impairments on energy detection based spectrum sensing for cognitive radio (CR) systems in multi-channel environments.
In particular, we provide closed-form expressions for the evaluation of the detection and false alarm probabilities, assuming Rayleigh fading.
Furthermore, we extend the analysis to the case of CR networks with cooperative sensing, where the secondary users suffer from different levels of RF imperfections, considering both scenarios of error free and imperfect reporting channel.
Numerical and simulation results demonstrate the accuracy of the analysis as well as the detrimental effects of RF imperfections on the spectrum sensing performance, which bring significant losses in the spectrum~utilization.
%Hence, the effects of hardware imperfections should be seriously taken into consideration when designing direct conversion CR receivers.

\end{abstract}


\section{Introduction}\label{S:Intro}

The rapid growth of wireless communications and the foreseen spectrum occupancy problems, due to the exponentially increasing consumer demands on mobile traffic and data, motivated the evolution of the concept of cognitive radio (CR) \cite{FCC_2002}.
CR systems require intelligent reconfigurable wireless devices, capable of sensing the conditions of the surrounding radio frequency (RF) environment and modifying their transmission parameters accordingly, in order to achieve the best overall performance, without interfering with other users \cite{ED_Bartlett}.
One fundamental task in CR is spectrum sensing, i.e., the identification of temporarily vacant portions of spectrum, over wide ranges of spectrum resources and determine the available spectrum holes on its own.
Spectrum sensing allows the exploitation of the under-utilized spectrum, which is considered to be an essential element in the operation of CRs.
Therefore, great amount of effort has been put to derive optimal, suboptimal, ad-hoc, and cooperative solutions to the spectrum sensing problem (see for example
\cite{
A_survey_of_spectrum_sensing_CR,
Opportunistic_Spectrum_Access_in_CR_Networks_Under_Imperfect_Spectrum_Sensing,
Relay_selection_in_CR_networks_with_interference_constrains,
Mutiuser_CR_networks_Joint_Impact_of_Direct_and_Relay_Communications,
%Optimal_Sensing_Transmission_Structure_for_Dynamic_Spectrum_Access,
A:SpecSensingOFM_CFO,
ED_Cooperative_Spectrum_Sensing_in_CR,
CR_GMD,
%A:Optical_multi_channel_cooperative_sensing_in_CR_networks,
A:EqualGainCombining_for_Coop_Spec_Sensing_in_CRs,
A:Optimization_of_cooperatice_spectrum_sensing_with_ED_in_CR_networks,
%A:Sensor_Selection_and_Optimal_Energy_Detection_Threshold_for_efficient_cooperative_spectrum_sensing,
A:On_the_performance_of_eigenvalue_based_Coop_Spect_Sensing_for_CR,
A:Unified_Analysis_of_Coop_Spect_Sensing_over_Composite_and_Generalized_Fading_Channels}).
However, the majority of these works ignore the imperfections associated with the RF front-end.
Such imperfections, which are encountered in the widely deployed low-cost direct-conversion radio (DCR) receivers (RXs), include in-phase (I) and quadrature-phase (Q) imbalance (IQI) \cite{Energy_Detection_under_IQI}, low-noise amplifier (LNA) nonlinearities \cite{CR_LNA} and phase noise (PHN)~\cite{ED_PHN}.

%\subsection{Related Work}
The effects of RF imperfections in general were studied in several works~\cite{
%Effects_of_IQI_on_blind_spectrum_sensing_for_OFDMA_overlay_CR,
B:Schenk-book,
B:wenk2010mimo,
RF_impairments_generalized_model,
A:A_new_look_at_dual_hop_relaying_performance_limits_with_hw_impairments,
Sensitivity_of_Spectrum_Sensing_to_RF_impairments,
Cyclostationary_Sensing_of_OFDM_RF_impairments,
IQI_IRR_practical_values,
A:IQI_TX_RX_AF_Alouini,
A:IQI_in_AF_Nakagami_m,
A:OFDM_OR_IQI,
A:IQI_in_Two_Way_AF_relaying,
A:Impairments_on_AF_relaying,
C:Massive_MIMO_systems_with_HW_constrained_BS,
A:Joint_Comp_IQI_and_PHN,
ED_PHN,
A:Joint_Mitigation_of_Nonlinear_and_baseband_distortion_in_wideband_DCRs,
C:High_Dynamic_Range_RF_FEs_from_multiband_multistandar_to_CR,
C:Implementation_issues_in_spectrum_sensing_for_CR,
Likelihood_based_specrum_sensing_of_OFDM_IQI,
Effects_of_IQI_on_blind_spectrum_sensing_for_OFDMA_overlay_CR
}.
However, only recently, the impacts of RF imperfections in the spectrum sensing
capabilities of CR was investigated
\cite{
C:High_Dynamic_Range_RF_FEs_from_multiband_multistandar_to_CR,
C:Implementation_issues_in_spectrum_sensing_for_CR,
Sensitivity_of_Spectrum_Sensing_to_RF_impairments,
Cyclostationary_Sensing_of_OFDM_RF_impairments,
Likelihood_based_specrum_sensing_of_OFDM_IQI,
Effects_of_IQI_on_blind_spectrum_sensing_for_OFDMA_overlay_CR,
Energy_Detection_under_IQI,
ED_PHN}.
In particular, the importance of improved front-end linearity and sensitivity was illustrated in
\cite{C:High_Dynamic_Range_RF_FEs_from_multiband_multistandar_to_CR} and
\cite{C:Implementation_issues_in_spectrum_sensing_for_CR},
while the impacts of RF impairments in DCRs on single-channel energy and/or cyclostationary based sensing were discussed in
\cite{Sensitivity_of_Spectrum_Sensing_to_RF_impairments} and
\cite{Cyclostationary_Sensing_of_OFDM_RF_impairments}.
Furthermore, in
\cite{Likelihood_based_specrum_sensing_of_OFDM_IQI} the authors presented closed-form expressions for the detection and false alarm probabilities for the Neyman-Pearson detector, considering the spectrum sensing problem in single-channel orthogonal frequency division multiplexing (OFDM) CR RX, under the joint effect of transmitter and receiver IQI.
On the other hand, multi-channel sensing under IQI was reported in
\cite{Effects_of_IQI_on_blind_spectrum_sensing_for_OFDMA_overlay_CR},
where a three-level hypothesis blind detector was introduced.
Moreover, the impact of RF IQI on energy detection (ED) for both single-channel
and multi-channel DCRs was investigated in
\cite{Energy_Detection_under_IQI}, where it was shown that the false alarm probability in a multi-channel environment increases significantly, compared to the ideal RF RX case. Additionally, in
\cite{ED_PHN}, the authors analyzed the effect of PHN on ED, considering a multi-channel DCR and additive white Gaussian noise (AWGN) channels, whereas in
\cite{A:Spectrum_Sensing_Under_RF_Non_linearities}, the impact of third-order non-linearities on the detection and false alarm probabilities for classical and cyclostationary energy detectors considering imperfect LNA, was~investigated.

In this work, we investigate the impact on the multi-channel energy-based spectrum sensing mechanism of the joint effects of several RF impairments, such as LNA non-linearities, PHN and IQI. After assuming
flat-fading Rayleigh channels and complex Gaussian primary user (PU) transmitted signals, and approximating the joint effects of RF impairments by a complex Gaussian process (an approximation which has been validated both in theory and by experiments, see
\cite{A:IQI_in_AF_Nakagami_m} and the references therein), we derive closed-form expressions for
the probabilities of false alarm and detection. Based on these expressions, we investigate the impact of RF impairments on ED. Specifically, the contribution of this paper can be summarized as~follows:
\begin{itemize}
\item We, first, derive analytical closed-form expressions for the false alarm and detection probabilities for an ideal RF front-end ED detector, assuming flat fading Rayleigh channels and complex Gaussian transmitted signals. To the best of the authors' knowledge, this is the first time that such expressions are presented in the open technical literature, under these~assumptions.
\item Next, a signal model that describes the joint effects of all RF impairments is presented. This model is built upon an approximation of the joint effects of RF impairments by a complex Gaussian process \cite{A:IQI_in_AF_Nakagami_m} and is tractable to algebraic manipulations.
\item Analytical closed-form expressions are provided for the evaluation of false alarm and detection probabilities of multi-channel EDs constrained by RF impairments, under Rayleigh fading. Based on this framework, the joint effects of RF impairments on spectrum sensing performance are~investigated.
\item Finally, we address an analytical study for the detection capabilities of cooperative spectrum sensing scenarios considering both cases of ideal ED detectors and multi-channel EDs constrained by RF impairments.
\end{itemize}

The remainder of the paper is organized as follows. The system and signal model for both ideal and hardware impaired RF front-ends are described in Section \ref{sec:SSM}.
The analytical framework for evaluating the false alarm and detection probabilities, when both ideal sensing or RF imperfections are considered,  are provided in Section \ref{sec:Probabilities}.
Moreover, analytical closed-form expression for deriving the false alarm and detection probabilities, when a cooperative spectrum sensing with decision fusion system is considered, are provided in Section \ref{sec:Cooperative_Spectrum_Sensing}.
Numerical and simulation results that illustrate the detrimental effects of RF impairments in spectrum sensing are presented in Section \ref{sec:Numerical_Results}. Finally, Section~\ref{sec:Conclusions} concludes the paper by summarizing our main findings.

\subsubsection*{Notations}
Unless otherwise stated, $(x)^{*}$ stands for the complex conjugate of $x$, whereas $\Re\left\{ x\right\} $ and $\Im\left\{ x\right\} $ represent the real and imaginary part of $x$, respectively.
The operators $E\left[\cdot\right]$ and $\left|\cdot\right|$ denote the statistical expectation and the absolute value, respectively.
The sign of a real number $x$ is returned by the operator $\sign\left(x\right)$.
The operator $\card\left(\mathcal{A}\right)$ returns the cardinality of the set $\mathcal{A}$.
%Additionally, $\{a_{k}\}_{k=1}^{n}$ is a shorthand notation for set $\left\{a_1, a_2, \cdots, a_n\right\}$.
$U\left(x\right)$ and $\exp\left(x\right)$ denote the unit step function and the exponential function, respectively.
The lower \cite[Eq. (8.350/1)]{B:Gra_Ryz_Book} and upper incomplete Gamma functions \cite[Eq. (8.350/2)]{B:Gra_Ryz_Book} are represented by $\gamma\left(\cdot,\cdot\right)$ and $\Gamma\left(\cdot,\cdot\right)$, respectively, while the Gamma function \cite[Eq. (8.310)]{B:Gra_Ryz_Book} is denoted by $\Gamma\left(\cdot\right)$.
Moreover, $\Gamma\left(a,x,b,\beta\right)=\int_{x}^{\infty}t^{a-1}\exp\left(-t-b t^{-\beta}\right) dt$ is the extended incomplete Gamma function defined by \cite[Eq. (6.2)]{B:chaudhry2001class}.
Finally, $\Q\left(x\right)=\frac{1}{\sqrt{2\pi}}\int_{x}^{\infty}\exp\left(-t^{2}/2\right)dt$ is the Gaussian Q-function.

\section{System and signal model}\label{sec:SSM}

In this section, we briefly present the ideal signal model, which is referred to as ideal RF front-end in what follows.
Build upon that, we demonstrate the practical signal model, where the RX is considered to suffer from RF imperfections, such as LNA nonlinearities, PHN and IQI.
Note that it is assumed that $K$ RF channels are down-converted to baseband using the wideband direct-conversion principle, which is referred to as multi-channel down-conversion~\cite{Direct_conversion}.

\subsection{Ideal RF front-end\label{sub:Ideal-RF-front-end}}

The two hypothesis, namely absence/presence of primary user (PU), is denoted with parameter $\theta_{k}\in\left\{ 0,1\right\}$. Suppose the $n$-th sample of the PU signal, $s\left(n\right),$ is conveyed over a flat-fading wireless channel, with channel gain, $h\left(n\right),$ and additive noise $w\left(n\right)$. The received wideband RF signal is passed through various RF front-end stages, including filtering, amplification, analog I/Q demodulation (down-conversion) to baseband and sampling. The wideband channel after sampling is assumed to have a bandwidth of $W$ and contain $K$ channels, each having bandwidth $W_{ch}=W_{sb}+W_{gb}$,
where $W_{sb}$ and $W_{gb}$ are the signal band and total guard band bandwidth within this channel, respectively. Additionally, it is assumed that the sampling is performed with rate $W$. Note, that the rate of the signal is reduced by a factor of $L=W/W_{sb}\geq K$, where for simplicity we assume $L\in\mathbb{Z}$.

Under the ideal RF front-end assumption, after the selection filter, the $n-$th sample of the baseband equivalent received signal vector for the $k^{\text{th}}$ channel
($k\in S\left\{ -K/2,\ldots,-1,1\ldots,K/2\right\} $) is given by
\begin{align}
r_{k}\left(n\right) & =\Re\left\{ r_{k}\left(n\right)\right\} +j\Im\left\{ r_{k}^{\text{}}\left(n\right)\right\}
%\\&
=\theta_{k}h_{k}\left(n\right)s_{k}\left(n\right)+w_{k}\left(n\right),\label{Rx_ideal_signal_model}
\end{align}
where $h_{k}$, $s_{k}$ and $w_{k}$ are zero-mean circular symmetric complex white Gaussian (CSCWG) processes with variances $\sigma_{h}^{2}$, $\sigma_{s}^{2}$ and $\sigma_{w}^{2}$, respectively. Furthermore,
\begin{align}
\Re & \left\{ r_{k}^{\text{}}\left(n\right)\right\} =\theta_{k}\Re\left\{ h_{k}\left(n\right)\right\} \Re\left\{ s_{k}\left(n\right)\right\}
%\nonumber \\&
-\theta_{k}\Im\left\{ h_{k}\left(n\right)\right\} \Im\left\{ s_{k}\left(n\right)\right\} +\Re\left\{ w_{k}\left(n\right)\right\} ,\\
\Im & \left\{ r_{k}^{\text{}}\left(n\right)\right\} =\theta_{k}\Im\left\{ h_{k}\left(n\right)\right\} \Re\left\{ s_{k}\left(n\right)\right\}
%\nonumber \\&
+\theta_{k}\Re\left\{ h_{k}\left(n\right)\right\} \Im\left\{ s_{k}\left(n\right)\right\} +\Im\left\{ w_{k}\left(n\right)\right\} .
\end{align}

\subsection{Non-ideal RF front-end\label{sub:Non-ideal-RF-front-end}}

%The direct-conversion wideband sensing scenario is very sensitive to RF circuit impairments, such as LNA nonlinearities, PHN and~IQI.
In the case of non-ideal RF front-end, the $n$-th sample of the impaired baseband equivalent received signal vector for the $k^{\text{th}}$channel is
given by~\cite{Energy_Detection_under_IQI} and~\cite{B:Schenk-book}
\begin{align}
r_{k}\left(n\right) & =\Re\left\{ r_{k}\left(n\right)\right\} +j\Im\left\{ r_{k}\left(n\right)\right\}
%\\&
=\xi_{k}\left(n\right)\theta_{k}h_{k}\left(n\right)s_{k}\left(n\right)+\eta_{k}\left(n\right)+w_{k}\left(n\right),\label{Rx_signal_model}
\end{align}
with
\begin{align}
% &
 \Re\left\{ r_{k}\left(n\right)\right\}\hspace{-0.1cm} =\hspace{-0.1cm}\theta_{k}\Re\left\{ h_{k}\left(n\right)\xi_{k}\right\} \Re\left\{ s_{k}\left(n\right)\right\}
 %\nonumber \\&
\hspace{-0.1cm} - \hspace{-0.1cm}
 \theta_{k}\Im\left\{ h_{k}\left(n\right)\xi_{k}\right\} \Im\left\{ s_{k}\left(n\right)\right\}
 \hspace{-0.1cm} + \hspace{-0.1cm}
 \Re\left\{ \eta_{k}\left(n\right)
 \hspace{-0.1cm}+\hspace{-0.1cm}
 w_{k}\left(n\right)\right\} ,
\end{align}
and
\begin{align}
% &
 \Im\left\{ r_{k}\left(n\right)\right\} \hspace{-0.1cm}=\hspace{-0.1cm}\theta_{k}\Im\left\{ h_{k}\left(n\right)\xi_{k}\right\} \Re\left\{ s_{k}\left(n\right)\right\}
 %\nonumber \\&
 \hspace{-0.1cm} - \hspace{-0.1cm}
 \theta_{k}\Re\left\{ h_{k}\left(n\right)\xi_{k}\right\} \Im\left\{ s_{k}\left(n\right)\right\}
 \hspace{-0.1cm} + \hspace{-0.1cm}
 \Im\left\{ \eta_{k}\left(n\right)
 \hspace{-0.1cm} + \hspace{-0.1cm}
 w_{k}\left(n\right)\right\} ,
\end{align}
where $\xi_{k}$ denotes the amplitude and phase rotation due to PHN caused by common phase error (CPE), LNA nonlinearities and IQI,
and is given by
\begin{equation}
\xi_{k}=\gamma_{0}K_{1}\alpha,\label{eq:ksi}
\end{equation}
while $\eta_{k}$ denotes the distortion noise from impairments in the RX, and specifically due to PHN caused by inter carrier interference (ICI), IQI and non-linear distortion noise, and is given by
\begin{align}
\eta_{k}\left(n\right) & =K_{1}\left(\gamma_{o}e_{k}\left(n\right)+\psi_{k}\left(n\right)\right)%\nonumber \\&
+K_{2}\left(\gamma_{o}^{*}\left(\alpha \theta_{-k} h_{-k}^{*}\left(n\right)s_{-k}^{*}\left(n\right)+e_{-k}^{*}\left(n\right)\right)\right)
%\nonumber \\&
 +K_{2}\psi_{-k}^{*}\left(n\right).
 \label{eta}
\end{align}
After denoting as $\Theta_{k}=\left\{ \theta_{k-1},\theta_{k+1}\right\} $
and $H_{k}=\left\{ h_{k-1},h_{k+1}\right\} $, this distortion noise
term can be modeled as
%\begin{equation}
$\eta_{k}\sim\mathcal{CN}\left(0,\sigma_{\eta_{k}}^{2}\right),$
%\end{equation}
with
\begin{align}
\sigma_{\eta_{k}}^{2}
%&
\hspace{-0.1cm}=\hspace{-0.1cm}
\left|\gamma_{0}\right|^{2}
\hspace{-0.1cm}
\left(\left|K_{1}\right|^{2}\sigma_{e,k}^{2}
\hspace{-0.1cm}+\hspace{-0.1cm}
\left|K_{2}\right|^{2}\sigma_{e,-k}^{2}\right)
%\nonumber \\
&
\hspace{-0.1cm}+\hspace{-0.1cm}
\left|K_{1}\right|^{2}\sigma_{\psi\left|H_{k},\Theta_{k}\right.}^{2}\hspace{-0.1cm}
\hspace{-0.1cm}+\hspace{-0.1cm}
\left|K_{2}\right|^{2}\sigma_{\psi\left|H_{-k},\Theta_{-k}\right.}^{2} \hspace{-0.1cm}
%\nonumber \\&
\hspace{-0.1cm}+\hspace{-0.1cm}\left|\gamma_{0}\right|^{2}\left|K_{2}\right|^{2}\left|\alpha\right|^{2}\theta_{-k}
\hspace{-0.1cm}
\left|h_{-k}\right|^{2}
\hspace{-0.1cm}
\sigma_{s}^{2}.\label{sigma_eta}
\end{align}
It should be noted that this model has been supported and validated
by many theoretical investigations and measurements \cite{MIMO_transmission_with_residual_transmit_RF_impairments,
A:IQI_in_AF_Nakagami_m,
C:Massive_MIMO_systems_with_HW_constrained_BS,
B:wenk2010mimo,
A_theoretical_characterization_of_nonlinear_distortion_effects_in_OFDM_systems,
Impairments_on_AF_relaying,
Experimental_Investigation_of_TDD_Reciprocity_Based_ZF,
RF_impairments_generalized_model}.

Next, we describe how the various parameters in (\ref{eq:ksi}), (\ref{eta}) and (\ref{sigma_eta}) stem from the imperfections associated with the RF front-end.

\subsubsection*{LNA Nonlinearities}

The parameters $\alpha$ and $e_{k}$ respresent the nonlinearity parameters, which model the amplitude/phase distortion and the nonlinear distortion noise, respectively.
According to Bussgang's theorem \cite{papoulis}, $e_{k}$ is a zero-mean Gaussian error term with variance $\sigma_{e_{k}}^{2}$. Considering an ideal clipping power amplifier (PA), the amplification
factor $\alpha$ and the variance $\sigma_{e_{k}}^{2}$, are given by
\begin{equation}
\alpha=1-\exp\left(-\IBO\right)+\sqrt{2\pi}\IBO\Q\left(2\IBO\right),
\end{equation}
\begin{equation}
\sigma_{e_{k}}^{2}=\sigma_{s}^{2}\left(1-\alpha^{2}-\exp\left(-\IBO\right)\right),
\end{equation}
where $\IBO=A_{o}^{2}/\sigma_{s}^{2}$ denotes the input back-off
factor and $A_{o}$ is the PA's clipping level.

Furthermore, if a polynomial model is employed to describe the effects of nonlinearities, the amplification factor $\alpha$ and the variance $\sigma_{e_{k}}$, are given by
\begin{align}
\alpha = \sum_{n=0}^{M-1}\beta_{n+1} 2^{-n/2} \sigma_{s}^{2} \Gamma\left(1+n/2\right), \\
\sigma_{e_{k}} = \sum_{n=2}^{2M}\gamma_{n} 2^{-n/2} \sigma_{s}^{2} \Gamma\left(1+n/2\right) - \left|a\right|^{2}\sigma_{s}^{2},
\end{align}
where
\begin{align}
\gamma_{n}=\sum_{m=1}^{n-1}\widehat{\beta}_{m}\widehat{\beta}_{n-m}^{*},
%\end{align}
\text{ and }
%\begin{align}
\widehat{\beta}_{m}=\left\{\begin{array}{c l}\beta_{m}, & 1\leq m \leq M+1 \\ 0, & m>M+1 \end{array} \right.
\end{align}

\subsubsection*{I/Q Imbalance}

The IQI coefficients $K_{1}$ and $K_{2}$ are given by
\begin{align}
K_{1}=\frac{1+\epsilon e^{-j\theta}}{2}\text{ and }K_{2}=\frac{1-\epsilon e^{j\theta}}{2},
\end{align}
with $\epsilon$ and $\theta$ denote the amplitude and phase mismatch, respectively. It is noted that for perfect I/Q matching, this imbalance parameters become $\epsilon=1$, $\theta=0$; thus in this case $K_{1}=1$ and $K_{2}=0$. The coefficients $K_{1}$ and $K_{2}$ are related through
$K_{1}=1-K_{2}^{*}$
and the image rejection ratio (IRR), which determines the amount of attenuation of the image frequency band, namely
$\rm{IRR}=\left|{K_{1}}/{K_{2}}\right|^{2}$.
With practical analog front-end electronics, $\rm{IRR}$ is typically in the range of $20-40$~$\rm{dB}$~\cite{Direct_conversion,IQI_IRR_practical_values,Boul1508:Effects}.

\subsubsection*{Phase noise}

The parameter, $\gamma_{0}$, stands for CPE, which is equal for all channels, and $\psi_{k}$ represents the ICI from all other neighboring channels due to spectral regrowth caused by PHN.
Notice that, since the typical $3\text{ dB}$ bandwidth values for the oscillator process is in the order of few tens or hundreds of Hz, with rapidly fading spectrum after this point (approximately $10\text{dB}/\text{decade}$), for channel bandwidth that is typical few tens or hundreds $\rm{KHz}$, the only effective interference is due to leakage from successive neighbors only \cite{ED_PHN}. Consequently, the ICI term can be approximated as \cite{ED_PHN}
\begin{align}
\psi_{k}\left(n\right) & \approx\theta_{k-1}\gamma\left(n\right)h_{k-1}\left(n\right)s_{k-1}\left(n\right)
%\nonumber \\&
+\theta_{k+1}\gamma\left(n\right)h_{k+1}\left(n\right)s_{k+1}\left(n\right),
\end{align}
with $\gamma\left(n\right)=\exp\left({j\phi\left(n\right)}\right)$ and $\phi\left(n\right)$ being a discrete Brownian error process, i.e.,
%\begin{align}
$\phi\left(n\right)=\sum_{m=1}^{n}\phi\left(m-1\right)+\epsilon\left(n\right),$
%\end{align}
where $\epsilon\left(n\right)$ is a zero mean real Gaussian variable with variance
%\begin{align}
$\sigma_{\epsilon}^{2}=\frac{4\pi\beta}{W}$
%\end{align}
and $\beta$ being the $3\text{ }\rm{dB}$ bandwidth of the local oscillator~process.

The interference term $\psi_{k}$ in \eqref{eta} might have zero or non-zero contribution depending on the existence of PU signals in the successive neighboring channels. In general, this term is typically non-white and strictly speaking cannot be modeled by a Gaussian process. However, for practical $3$ $\rm{dB}$ bandwidth of the oscillator process, the influence of the regarded impairments can all be modeled as a zero-mean Gaussian process with $\sigma_{\left.\psi_{k}\right|\left\{ H_{k},\Theta_{k}\right\} }^{2}$ given by
\begin{align}
\sigma_{\left.\psi\right|\left\{ H_{k},\Theta_{k}\right\} }^{2}
%&
=\theta_{k-1}A_{k-1}\left|h_{k-1}\left(n\right)\right|^{2}\sigma_{s}^{2}
%\nonumber \\&
+\theta_{k+1}A_{k+1}\left|h_{k+1}\left(n\right)\right|^{2}\sigma_{s}^{2},
\label{sigma_psi}
\end{align}
where
\begin{equation}
A_{k-1}=\frac{\left|I\left(f_{k-1}-f_{k}+f_{\text{cut-off}}\right)-I\left(f_{k-1}-f_{k}-f_{\text{cut-off}}\right)\right|}{2\pi f_{\text{cut-off}}},
\label{A_kplyn1}
\end{equation}
\begin{equation}
A_{k+1}=\frac{\left|I\left(f_{k+1}-f_{k}+f_{\text{cut-off}}\right)-I\left(f_{k+1}-f_{k}-f_{\text{cut-off}}\right)\right|}{2\pi f_{\text{cut-off}}},
\label{A_ksyn1}
\end{equation}
and $f_{k}$ is the centered normalized frequency of the $k^{th}$ channel, i.e.,
%\begin{align}
$f_{k}=\sign\left(k\right)\frac{2\left|k\right|-1}{2K}$
%\end{align}
and $f_{\text{cut-off}}$ is the normalized cut-off frequency of the $k^{th}$ channel, which can be obtained~by
%\begin{align}
$f_{\text{cut-off}}=\frac{W_{sb}}{2W}.$
%\end{align}
Furthermore,
\begin{align}
I\left(f\right)= &
\left(f_{\text{cut-off}}-f\right)\tan^{-1}\hspace{-0.1cm}\left(\delta\tan\left(\pi\left(f_{\text{cut-off}}-f\right)\right)\right)
%\nonumber \\&
+\left(f_{\text{cut-off}}+f\right)\tan^{-1}\hspace{-0.1cm}\left(\delta\tan\left(-\pi\left(f_{\text{cut-off}}+f\right)\right)\right)
\nonumber \\ &
- \frac{1}{\delta}\left(\left(f_{\text{cut-off}}+f\right)\cot\left(\pi\left(f_{\text{cut-off}}+f\right)\right)\right.
%\nonumber \\&
-\left.\left(f_{\text{cut-off}}-f\right)\cot\left(\pi\left(f_{\text{cut-off}}-f\right)\right)\right)\nonumber \\&
+\frac{1}{\pi\delta}\left(\log\left(\left|sin\left(\pi\left(f_{\text{cut-off}}+f\right)\right)\right|\right)\right.
% \nonumber \\&
 +\left.\log\left(\left|sin\left(\pi\left(f_{\text{cut-off}}-f\right)\right)\right|\right)\right),
%
%B\left(f\right)-C\left(f\right),
\end{align}
%where $B\left(f\right)$ and $C\left(f\right)$ are auxiliary functions, given by
%\begin{align}
%B\hspace{-0.1cm}\left(f\right) &\hspace{-0.1cm} = \hspace{-0.1cm}\left(f_{\text{cut-off}}-f\right)\tan^{-1}\hspace{-0.1cm}\left(\delta\tan\left(\pi\left(f_{\text{cut-off}}-f\right)\right)\right)
%\nonumber \\&
%+\left(f_{\text{cut-off}}+f\right)\tan^{-1}\hspace{-0.1cm}\left(\delta\tan\left(-\pi\left(f_{\text{cut-off}}+f\right)\right)\right),\label{B}
%\end{align}
%and
%\begin{align}
%C\left(f\right) & =\frac{1}{\delta}\left(\left(f_{\text{cut-off}}+f\right)\cot\left(\pi\left(f_{\text{cut-off}}+f\right)\right)\right.
%\nonumber \\&
%-\left.\left(f_{\text{cut-off}}-f\right)\cot\left(\pi\left(f_{\text{cut-off}}-f\right)\right)\right)\nonumber \\
% & -\frac{1}{\pi\delta}\left(\log\left(\left|sin\left(\pi\left(f_{\text{cut-off}}+f\right)\right)\right|\right)\right.
% \nonumber \\&
% +\left.\log\left(\left|sin\left(\pi\left(f_{\text{cut-off}}-f\right)\right)\right|\right)\right),\label{C}
%\end{align}
%respectively,
with
%\begin{equation}
$\delta=\frac{\exp\left({-2\pi\beta/W}\right)+1}{\exp\left({-2\pi\beta/W}\right)-1}.$
%\end{equation}
Due to Eqs. (\ref{A_kplyn1}) and (\ref{A_ksyn1}), it follows that
$A_{k-1}=A_{k+1}$.

\subsubsection*{Joint effect of RF impairments}
\begin{figure}
\centering\includegraphics[width=0.7\linewidth,trim=0 0 0 0,clip=false]{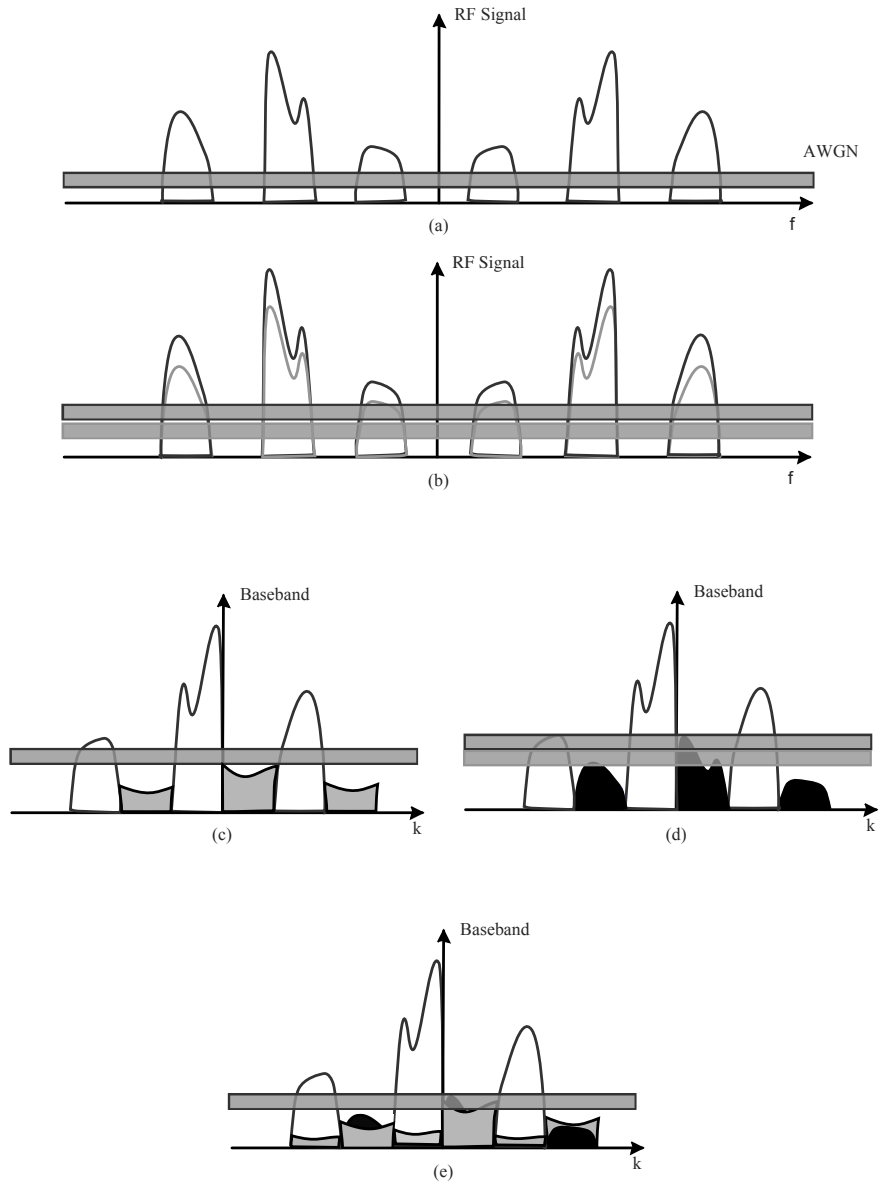}
\caption{Spectra of the received signal:
(a) before LNA (passband RF signal),
(b) after LNA (passband RF signal),
(c) after down-conversion (baseband signal), when local oscillator's PHN is considered to be the only RF imperfection,
(d) after down-conversion (baseband signal), when IQI is considered to be the only RF imperfection, (e) after down-conversion (baseband signal), the joint effect of LNA nonlinearities, PHN and IQI.}
\label{fig:RF_imp_effects}
\end{figure}

Here, we explain the joint impact of RF imperfections in the spectra of the down-converted received signal.  Comparing Eq. \eqref{Rx_signal_model} with Eq. \eqref{Rx_ideal_signal_model}, we observe that the RF imperfections result to not only amplitude/phase distortion, but also neighbor and mirror interference, as demonstrated intuitively in Fig.~\ref{fig:RF_imp_effects}.

According to \eqref{eq:ksi} and \eqref{sigma_eta}, LNA nonlinearities cause amplitude/phase distortion and an additive nonlinear distortion noise, whereas, based on \eqref{sigma_psi}, PHN causes interference to the received baseband signal at the $k^\text{th}$ channel, due to the received baseband signals at the neighbor channels $k-1$ and $k+1$.
%Furthermore, IQI results to interference to the received baseband signal at the $k^\text{th}$ channel by the image signal at the channel $-k$.

Moreover, based on (\ref{sigma_eta}), the joint effects of PHN and IQI, described by the terms $\left|K_{1}\right|^{2}\sigma_{\psi\left|H_{k},\Theta_{k}\right.}^{2}$, $\left|K_{2}\right|^{2}\sigma_{\psi\left|H_{-k},\Theta_{-k}\right.}^{2}$ and $\left|\gamma_{0}\right|^{2}\left|K_{2}\right|^{2}\left|\alpha\right|^{2}\theta_{-k}\left|h_{-k}\right|^{2}\sigma_{s}^{2}$, result to interference to the signal at the $k^\text{th}$  ($k\in\{-\frac{K}{2}+1,\cdots,\frac{K}{2}+1\}$) channel by the signals at the channels $-k-1$, $-k$, $-k+1$, $k-1$ and $k+1$.
Note that if $k=-\frac{K}{2}$ or $k=\frac{K}{2}$, then PHN and IQI cause interference to the signal at the $k^\text{th}$ channel due to the signals at the channels $-k$, $-k+1$ and $k-1$.
Consequently, in this case, the terms that refer to the signals at the channels $-k-1$ and $k+1$ should be omitted.

Furthermore, the joint effects of LNA nonlinearties and IQI are described by the first term  and the last terms in \eqref{sigma_eta}, i.e., $\left|K_{1}\right|^{2}\sigma_{e,k}^{2}+\left|K_{2}\right|^{2}\sigma_{e,-k}^{2}$ and $\left|\gamma_{0}\right|^{2}\left|K_{2}\right|^{2}\left|\alpha\right|^{2}\theta_{-k}\left|h_{-k}\right|^{2}\sigma_{s}^{2}$, respectively, and result to additive distortion noises and mirror channel interference.
Finally, the amplitude and phase distortion caused by the joint effects of all RF imperfections are modeled by the parameter $\xi$ described in~\eqref{eq:ksi}.

\section{False Alarm/Detection Probabilities for Channel Detection\label{sec:Probabilities}}

In the classical ED, the energy of the received signals is used to determine whether a channel is idle or busy. Based on the signal model described in Section \ref{sec:SSM}, the ED calculates the test statistics for the $k$ channel as
\begin{align}
T_{k} & =\frac{1}{N_{s}}\sum_{m=0}^{N_{s}-1}\left|r_{k}\left(n\right)\right|^{2}
%\\&
=\frac{1}{N_{s}}\sum_{m=0}^{N_{s}-1}\Re\left\{ r_{k}\left(n\right)\right\} ^{2}+\Im\left\{ r_{k}\left(n\right)\right\} ^{2},\label{ED_classic}
\end{align}
where $N_{s}$ is the number of complex samples used for sensing. This test statistic is compared against a threshold $\gamma_{th}\left(k\right)$ to yield the sensing decision,
i.e., the ED decides that the channel $k$ is busy if $T_{k}>\gamma_{th}\left(k\right)$ or idle~otherwise.

\subsection{Ideal RF front-end}
Based on the signal model presented in \ref{sub:Ideal-RF-front-end}
and taking into consideration that
\begin{align}
\sigma^{2} & =E\left[\Re\left\{ r_{k}\right\} ^{2}\right]=E\left[\Im\left\{ r_{k}\right\} ^{2}\right]
%\nonumber \\&
=\theta_{k}\left(\Re\left\{ h_{k}\right\} ^{2}+\Im\left\{ h_{k}\right\} ^{2}\right)\frac{\sigma_{s}^{2}}{2}+\frac{\sigma_{w}^{2}}{2},
\end{align}
and $E\left[\Re\left\{ r_{k}\right\} \Im\left\{ r_{k}\right\} \right]=0$
for a given channel realization $h_{k}$ and channel occupation $\theta_{k}$,
the received energy follows chi-square distribution with $2N_{s}$
degrees of freedom and cumulative distribution function (CDF) given
by
\begin{equation}
F_{T_{k}}\left(x\left|h_{k},\theta_{k}\right.\right)=\frac{\gamma\left(N_{s},\frac{N_{s}x}{2\sigma^{2}}\right)}{\Gamma\left(N_{s}\right)}.\label{eq:ideal_cond}
\end{equation}

The following theorem returns a closed-form expression for the CDF
of the test statistics assuming that the channel is busy.
\begin{thm}
\label{thm:CDF_ideal}The CDF of the energy statistics assuming an ideal RF front end and a busy channel can be evaluated by
\begin{align}
 & F_{T_{k}}\left(x\left|\theta_{k}=1\right.\right)=
 1- \exp\left(\frac{\sigma_{w}^2}{\sigma_h^2 \sigma_s^2}\right)
 \sum_{k=0}^{N_s-1}\frac{1}{k!} \left(\frac{N_s x}{\sigma_h^2 \sigma_s^2}\right)^k \Gamma\left(-k+1,\frac{\sigma_w^2}{\sigma_h^2 \sigma_s^2},\frac{N_s x}{\sigma_h^2 \sigma_s^2},1\right),
%%%
%& F_{T_{k}}\left(x\left|\theta_{k}=1\right.\right)=
% 1- \frac{\exp\left(\frac{\sigma_{w}^2}{\sigma_h^2 \sigma_s^2}\right)}{\Gamma\left(N_s\right)} \sum_{k=0}^{N_s-1}\frac{\left(N_s-1\right)!}{k!} \left(\frac{N_s x}{\sigma_h^2 \sigma_s^2}\right)^k \Gamma\left(-k+1,\frac{\sigma_w^2}{\sigma_h^2 \sigma_s^2},\frac{N_s x}{\sigma_h^2 \sigma_s^2},1\right).
\label{CDF_theta_1_ideal}
\end{align}
%where $\Gamma\left(\cdot, \cdot, \cdot, \cdot\right)$ is the extended incomplete gamma function, which was described in the notation~paragraph.
\end{thm}
\begin{IEEEproof}
Since $h_{k}\sim\mathcal{CN}\left(0,\sigma_{h}^{2}\right)$, it follows that the parameter $\sigma^{2}$ follows exponential distribution with probability density function (PDF) given~by
\begin{equation}
f_{\sigma^{2}}\left(x\left|\theta_{k}=1\right.\right)=\frac{2\exp\left(\frac{\sigma_{w}^{2}}{\sigma_{s}^{2}\sigma_{h}^{2}}\right)}{\sigma_{s}^{2}\sigma_{h}^{2}}\exp\left(-\frac{2x}{\sigma_{s}^{2}\sigma_{h}^{2}}\right),
\end{equation}
with $x\in\left[\frac{\sigma_{w}^{2}}{2},\infty\right)$. Hence, the unconditional CDF can be expressed as
\begin{align}
F_{T_{k}}\left(x\left|\theta_{k}=1\right.\right)
%&
=\frac{1}{\Gamma\left(N_{s}\right)}\frac{2\exp\left(\frac{\sigma_{w}^{2}}{\sigma_{s}^{2}\sigma_{h}^{2}}\right)}{\sigma_{s}^{2}\sigma_{h}^{2}}
%\nonumber \\&
\int_{\frac{\sigma_{w}^{2}}{2}}^{\infty}\gamma\left(N_{s},\frac{N_{s}x}{2y}\right)\exp\left(-\frac{2y}{\sigma_{h}^{2}\sigma_{s}^{2}}\right)dy,
\end{align}
which is equivalent to
\begin{align}
 F_{T_{k}}\left(x\left|\theta_{k}=1\right.\right)
&
=\frac{1}{\Gamma\left(N_{s}\right)}\frac{2\exp\left(\frac{\sigma_{w}^{2}}{\sigma_{s}^{2}\sigma_{h}^{2}}\right)}{\sigma_{s}^{2}\sigma_{h}^{2}}
%\nonumber \\&
\int_{\frac{\sigma_{w}^{2}}{2}}^{\infty}\Gamma\left(N_{s}\right)\exp\left(-\frac{2y}{\sigma_{h}^{2}\sigma_{s}^{2}}\right)dy
\nonumber \\ &
- \frac{1}{\Gamma\left(N_{s}\right)}\frac{2\exp\left(\frac{\sigma_{w}^{2}}{\sigma_{s}^{2}\sigma_{h}^{2}}\right)}{\sigma_{s}^{2}\sigma_{h}^{2}}
%\nonumber \\&
\int_{\frac{\sigma_{w}^{2}}{2}}^{\infty}\Gamma\left(N_{s},\frac{N_{s}x}{2y}\right)\exp\left(-\frac{2y}{\sigma_{h}^{2}\sigma_{s}^{2}}\right)dy,
\end{align}
or
\begin{align}
F_{T_{k}}\left(x\left|\theta_{k}=1\right.\right)
%&
= 1 - \frac{1}{\Gamma\left(N_{s}\right)}\frac{2\exp\left(\frac{\sigma_{w}^{2}}{\sigma_{s}^{2}\sigma_{h}^{2}}\right)}{\sigma_{s}^{2}\sigma_{h}^{2}}
%\nonumber \\&
\int_{\frac{\sigma_{w}^{2}}{2}}^{\infty}\Gamma\left(N_{s},\frac{N_{s}x}{2y}\right)\exp\left(-\frac{2y}{\sigma_{h}^{2}\sigma_{s}^{2}}\right)dy .
 \label{Eq:F_T_k_th_1_proof}
\end{align}
Since $N_s$ is a positive integer, the upper incomplete Gamma function can be written as a finite sum \cite[Eq. (8.352/2)]{B:Gra_Ryz_Book}, and hence \eqref{Eq:F_T_k_th_1_proof} can be re-written as
%\begin{align}
%\Gamma\left(N_{s},\frac{N_{s} x}{2y} \right)= \sum_{k=0}^{N_{s}-1}\frac{\left(N_{s}-1\right)!}{k!} \left(\frac{N_{s} x}{2 y}\right)^{k} \exp\left(-\frac{N_{s} x}{2 y}\right).
%\label{Eq:Gamma_incomplete_Ns}
%\end{align}
%Substituting \eqref{Eq:Gamma_incomplete_Ns} to \eqref{Eq:F_T_k_th_1_proof}, yields
\begin{align}
F_{T_{k}}\left(x\left|\theta_{k}=1\right.\right)
%&
= 1 - \frac{2\exp\left(\frac{\sigma_{w}^{2}}{\sigma_{s}^{2}\sigma_{h}^{2}}\right)}{\sigma_{s}^{2}\sigma_{h}^{2}}
%\nonumber \\&
\sum_{k=0}^{N_{s}-1} \int_{\frac{\sigma_{w}^{2}}{2}}^{\infty}\frac{1}{k!} \left(\frac{N_{s} x}{2 y}\right)^{k} \exp\left(-\frac{N_{s} x}{2 y}-\frac{2y}{\sigma_{h}^{2}\sigma_{s}^{2}}\right)dy .
\label{Eq:F_T_k_th_2_proof}
\end{align}

After some algebraic manipulations and using \cite[Eq. (6.2)]{B:chaudhry2001class}, (\ref{Eq:F_T_k_th_2_proof}) can be written as in (\ref{CDF_theta_1_ideal}). This concludes
the proof.
\end{IEEEproof}

Based on the above analysis, the false alarm probability for the ideal RX can be obtained~by
\begin{align}
{\cal P}_{fa}(\gamma) & = P_r\left(T_{k}>\gamma\left|\theta_{k}=0 \right.\right)
=\frac{\Gamma\left(N_{s},\frac{N_{s}\gamma}{\sigma_{w}^{2}}\right)}{\Gamma\left(N_{s}\right)},
\label{Eq:P_FA_Ideal_RF}
\end{align}
while the probability of detection can be calculated~as
\begin{align}
{\cal P}_{d}(\gamma)& \hspace{-0.1cm}=\hspace{-0.1cm} P_r\left(T_{k}\hspace{-0.1cm} >\hspace{-0.1cm} \gamma\left|\theta_{k}\hspace{-0.1cm}=\hspace{-0.1cm}1 \right.\right)
%\nonumber \\ &
\hspace{-0.1cm}=\hspace{-0.1cm}{\exp\left(\frac{\sigma_{w}^2}{\sigma_h^2 \sigma_s^2}\right)} \hspace{-0.1cm}  \sum_{k=0}^{N_s-1}\hspace{-0.1cm}\frac{1}{k!} \hspace{-0.1cm}  \left(\frac{N_s \gamma}{\sigma_h^2 \sigma_s^2}\right)^k  \Gamma\left(-k+1,\frac{\sigma_w^2}{\sigma_h^2 \sigma_s^2},\frac{N_s \gamma}{\sigma_h^2 \sigma_s^2},1\right).
%%
%{\cal P}_{d}(\gamma)& \hspace{-0.1cm}=\hspace{-0.1cm} P_r\left(T_{k}\hspace{-0.1cm} >\hspace{-0.1cm} \gamma\left|\theta_{k}\hspace{-0.1cm}=\hspace{-0.1cm}1 \right.\right)
%\nonumber \\ &
%\hspace{-0.1cm}=\hspace{-0.1cm}\frac{\exp\left(\frac{\sigma_{w}^2}{\sigma_h^2 \sigma_s^2}\right)}{\Gamma\left(N_s\right)} \hspace{-0.1cm}  \sum_{k=0}^{N_s-1}\hspace{-0.1cm}\frac{\left(N_s-1\right)!}{k!} \hspace{-0.1cm}  \left(\frac{N_s \gamma}{\sigma_h^2 \sigma_s^2}\right)^k  \Gamma\left(-k+1,\frac{\sigma_w^2}{\sigma_h^2 \sigma_s^2},\frac{N_s \gamma}{\sigma_h^2 \sigma_s^2},1\right).
\label{Eq:P_D_Ideal_RF}
\end{align}

\subsection{Non-Ideal RF Front-End}

Based on the signal model presented in \ref{sub:Non-ideal-RF-front-end}, and assuming given channel realization and channel occupancy vectors
$H=\left\{ H_{-k},h_{-k},h_{k},H_{k}\right\} $ and $\Theta=\left\{ \Theta_{-k},\theta_{-k},\theta_{k},\Theta_{k}\right\} $,
respectively, it holds that
\begin{align}
\sigma^{2} &\hspace{-0.1cm}  = \hspace{-0.1cm} E\left[\Re\left\{ r_{k}\right\} ^{2}\right]\hspace{-0.1cm} = \hspace{-0.1cm} E\left[\Im\left\{ r_{k}\right\} ^{2}\right]
%\nonumber \\&
\hspace{-0.1cm} = \hspace{-0.1cm}  \theta_{k}\left(\Re\left\{ h_{k}\right\}^{2} \hspace{-0.1cm} + \hspace{-0.1cm} \Im\left\{ h_{k}\right\}^{2}\right)\hspace{-0.1cm} \left(\Re\left\{ \xi_{k}\right\} ^{2} \hspace{-0.1cm} + \hspace{-0.1cm} \Im\left\{ \xi_{k}\right\} ^{2}\right)\frac{\sigma_{s}^{2}}{2}
 %\nonumber \\&
 \hspace{-0.1cm} + \hspace{-0.1cm} \frac{\sigma_{w}^{2}\hspace{-0.1cm} +\hspace{-0.1cm} \sigma_{\eta_{k}}^{2}}{2},\label{sigma_received_non_ideal}
\end{align}
and $\Re\left\{ r_{k}\right\}$, $\Im\left\{ r_{k}\right\}$ are uncorrelated random variables, i.e., $E\left[\Re\left\{ r_{k}\right\} \Im\left\{ r_{k}\right\} \right]=0$.
Thus, the received energy, given by (\ref{ED_classic}), follows chi-square distribution with $2N_{s}$ degrees of freedom and CDF given by
\begin{align}
F_{T_{k}}\left(x\left|H,\Theta\right.\right)=\frac{\gamma\left(N_{s},\frac{N_{s}x}{2\sigma^{2}}\right)}{\Gamma\left(N_{s}\right)},
\label{Eq:F_Tk_cond_H_Theta}
\end{align}
where $\sigma^{2}$ can be expressed, after taking into account (\ref{sigma_eta}), (\ref{sigma_psi})
and (\ref{sigma_received_non_ideal}), as
\begin{align}
  \sigma^{2}
  &
  =\theta_{k}\mathcal{A}_{1}\left|h_{k}\right|^{2}+\theta_{k-1}\mathcal{A}_{2}\left|h_{k-1}\right|^{2}+\theta_{k+1}\mathcal{A}_{2}\left|h_{k+1}\right|^{2}
% &
 +\theta_{-k+1}\mathcal{A}_{3}\left|h_{-k+1}\right|^{2}
\nonumber \\ &
 +\theta_{-k-1}\mathcal{A}_{3}\left|h_{-k-1}\right|^{2}+\theta_{-k}\mathcal{A}_{4}\left|h_{-k}\right|^{2}
% \nonumber \\&
 +\mathcal{A}_{5}.\label{eq:sigma}
\end{align}
In the above equation,
%\begin{align}
$\mathcal{A}_{1}  =\left|\xi_{k}\right|^{2}\frac{\sigma_{s}^{2}}{2},$
%\\
$\mathcal{A}_{2}  =\left|K_{1}\right|^{2}A_{k-1}\frac{\sigma_{s}^{2}}{2},$
%\\
$\mathcal{A}_{3}  =\left|K_{2}\right|^{2}A_{-k+1}\frac{\sigma_{s}^{2}}{2},$
%\\
$\mathcal{A}_{4}  =\left|\gamma_{0}\right|^{2}\left|K_{2}\right|^{2}\left|a\right|^{2}\frac{\sigma_{s}^{2}}{2},$
%\end{align}
and
%\begin{align}
$\mathcal{A}_{5}=\frac{\sigma_{w}^{2}}{2}+\frac{\left|\gamma_{0}\right|^{2}}{2}\left(\left|K_{1}\right|^{2}\sigma_{e,k}^{2}+\left|K_{2}\right|^{2}\sigma_{e,-k}^{2}\right)$
%\end{align}
model the amplitude distortion due to the joint effects of RF impairments, the interference from the $k-1$ and $k+1$ channels, the interference from the $-k-1$ and $-k+1$ channels due to PHN, the mirror interference due to IQI, and the distortion noise due to the joint effects of RF impairments, respectively.

The following theorems return analytical closed-form expressions for the CDF of the energy test statistics for a given channel occupancy vector, when at least one channel of $\{-k-1,-k,-k+1,k-1,k,k+1\}$ is busy and when all channels are idle.
\begin{thm}
\label{thm:The-CDF-of-non-ideal}The CDF of the energy statistics
assuming an non-ideal RF front end and an arbitrary channel occupancy vector
$\Theta$ that is different than the all idle vector, can be evaluated by \eqref{Eq:F_Tk_final}, given at the top of the next page,
\begin{figure*}
\begin{align}
& F_{T_{k}}  \left(x\left|\Theta\right.\right)=\sum_{i=2}^{3}U\left(m_{i}-2\right)w_{1,i}w_{2,i}\mathcal{A}_{i} \exp\left(-\frac{\mathcal{A}_5}{\mathcal{A}_{i}}\right)
%\nonumber \\&
+\sum_{i=1}^{4}U\left(m_{i}-2\right)w_{1,i} \mathcal{A}_{i} \left(\mathcal{A}_{5}+\mathcal{A}_{i}\right) \exp\left(-\frac{\mathcal{A}_{5}}{\mathcal{A}_{i}}\right)
\nonumber \\
 & +\sum_{i=1}^{4}U\left(m_{i}-1\right)\left(U\left(1-m_{i}\right)-\mathcal{A}_{5}U\left(m_{i}-2\right)\right)w_{1,i}\mathcal{A}_{i} \exp\left(-\frac{\mathcal{A}_5}{\mathcal{A}_{i}}\right)
 \nonumber \\ &
- \sum_{i=2}^{3} \sum_{k=0}^{N_{s}-1} U\left(m_{i}-2\right) \frac{1}{k!} \frac{w_{1,i}w_{2,i}}{\mathcal{A}_{i}^{k-1}} \left(\frac{N_{s} x}{2}\right)^{k}  {\Gamma\left(-k+1,\frac{\mathcal{A}_{5}}{\mathcal{A}_{i}},\frac{N_s x}{2\mathcal{A}_{i}},1\right)}
 \nonumber \\
& -\sum_{i=1}^{4}\sum_{k=0}^{N_{s}-1}U\left(m_{i}-1\right)\left(U\left(1-m_{i}\right)-\mathcal{A}_{5}U\left(m_{i}-2\right)\right)\frac{1}{k!}\frac{w_{1,i}}{\mathcal{A}_{i}^{k-1}}\left(\frac{N_{s} x}{2}\right)^{k}
{\Gamma\left(-k+1,\frac{\mathcal{A}_{5}}{\mathcal{A}_{i}},\frac{N_s x}{2\mathcal{A}_{i}},1\right)}
\nonumber \\ &
 - \sum_{i=1}^{4}\sum_{k=0}^{N_{s}-1}U\left(m_{i}-2\right)
   \frac{1}{k!} \frac{w_{1,i}}{\mathcal{A}_{i}^{k-1}} \left(\frac{N_{s}x}{2}\right)^{k} {\Gamma\left(-k+2,\frac{\mathcal{A}_{5}}{\mathcal{A}_{i}},\frac{N_{s} x}{2\mathcal{A}_{i}},1\right)}.
 \label{Eq:F_Tk_final}
\end{align}
\hrulefill{}\vspace{-2pt}
\end{figure*}
where $w_{1,i}$ and $w_{2,i}$ are given by
\begin{align}
w_{1,i} & =\frac{\exp\left(\frac{\mathcal{A}_{5}}{\mathcal{A}_{i}}\right)}{\Gamma\left(m_{i}\right)\left(\prod_{j=1}^{4}\mathcal{A}_{j}^{m_{j}}\right)}\prod_{j=1,j\neq i}^{4}\left(\frac{1}{\mathcal{A}_{j}}-\frac{1}{\mathcal{A}_{i}}\right)^{-m_{j}},\label{eq:w1}
\end{align}
and
\begin{equation}
w_{2,i}=\sum_{j=1,j\neq i}m_{j}\left(\frac{1}{\mathcal{A}_{j}}-\frac{1}{\mathcal{A}_{i}}\right)^{-1},
\label{eq:w2}
\end{equation}
respectively.
\end{thm}
\begin{IEEEproof}
According to \cite{A:Karagiannidis-2006-ID448} and after some basic algebraic manipulations, its PDF can be written~as
\begin{align}
  f_{\sigma^{2}}\left(x\left|\Theta\right.\right)&=\sum_{i=2}^{3}U\left(m_{i}-2\right)w_{1,i}w_{2,i}\exp\left(-\frac{x}{\mathcal{A}_{i}}\right)\nonumber \\
 &+ \sum_{i=1}^{4}U\left(m_{i}-1\right)\left(U\left(1-m_{i}\right)-\mathcal{A}_{5}U\left(m_{i}-2\right)\right)w_{1,i}\exp\left(-\frac{x}{\mathcal{A}_{i}}\right)\nonumber \\
 & +\sum_{i=1}^{4}U\left(m_{i}-1\right)U\left(m_{i}-2\right)w_{1,i}x\exp\left(-\frac{x}{\mathcal{A}_{i}}\right),
 \label{Eq:Conditional_PDF_of_sigma}
\end{align}
where  $x\in\left[{\cal A}_{5},\infty\right)$, $m=\left[\theta_{k},\theta_{k-1}+\theta_{k+1},\theta_{-k+1}+\theta_{-k-1},\theta_{-k}\right]$,
$w_{1,i}$ and $w_{2,i}$ are defined by (\ref{eq:w1}) and (\ref{eq:w2})
respectively.

Based on the above, the CDF of the received energy, in case of non-ideal RF front-end, unconditioned with respect to $\Theta$, can be expressed~as
\begin{align}
F_{T_{k}}  \left(x\left|\Theta\right.\right)=\sum_{i=2}^{3}U\left(m_{i}-2\right)w_{1,i}w_{2,i}\mathcal{I}_{1,i}
%\nonumber \\&
&+\sum_{i=1}^{4}U\left(m_{i}-1\right)\left(U\left(1-m_{i}\right)-\mathcal{A}_{5}U\left(m_{i}-2\right)\right)w_{1,i}\mathcal{I}_{1,i}\nonumber \\
 & +\sum_{i=1}^{4}U\left(m_{i}-1\right)U\left(m_{i}-2\right)w_{1,i}\mathcal{I}_{2,i},\label{Eq:test1}
\end{align}
with
\begin{align}
\mathcal{I}_{1,i} & =\frac{1}{\Gamma\left(N_{s}\right)}\int_{\mathcal{A}_{5}}^{\infty}\exp\left(-\frac{y}{\mathcal{A}_{i}}\right)\gamma\left(N_{s},\frac{N_{s}x}{2y}\right)dy,\label{I1}\\
\mathcal{I}_{2,i} & =\frac{1}{\Gamma\left(N_{s}\right)}\int_{\mathcal{A}_{5}}^{\infty}y\exp\left(-\frac{y}{\mathcal{A}_{i}}\right)\gamma\left(N_{s},\frac{N_{s}x}{2y}\right)dy.\label{I2}
\end{align}
Eqs. (\ref{I1}) and (\ref{I2}), after some basic algebraic manipulations,
and using \cite[Eq. (8.352/2)]{B:Gra_Ryz_Book} and \cite[Eq. (6.2)]{B:chaudhry2001class}, can be written as
\begin{align}
\mathcal{I}_{1,i} & =\mathcal{A}_{i} \exp\left(-\frac{\mathcal{A}_5}{\mathcal{A}_{i}}\right)- \sum_{k=0}^{N_{s}-1}\frac{\left(N_{s}-1\right)!}{k!}\left(\frac{N_{s} x}{2}\right)^{k} \frac{1}{\mathcal{A}_{i}^{k+1}} \frac{\Gamma\left(-k+1,\frac{\mathcal{A}_{5}}{\mathcal{A}_{i}},\frac{N_s x}{2\mathcal{A}_{i}},1\right)}{\Gamma\left(N_{s}\right)},
\label{I1i_final}
\end{align}
and
\begin{align}
\mathcal{I}_{2,i} & \hspace{-0.1cm} = \hspace{-0.1cm}
\mathcal{A}_{i} \left(\mathcal{A}_{5}+\mathcal{A}_{i}\right) \exp\left(-\frac{\mathcal{A}_{5}}{\mathcal{A}_{i}}\right)
\hspace{-0.1cm}-\hspace{-0.1cm} \sum_{k=0}^{N_{s}-1}\hspace{-0.1cm} \frac{\left(N_{s}-1\right)!}{k!} \left(\frac{N_{s}x}{2}\right)^{k} \frac{1}{\mathcal{A}_{i}^{k+1}}\frac{\Gamma\left(-k+2,\frac{\mathcal{A}_{5}}{\mathcal{A}_{i}},\frac{N_{s} x}{2\mathcal{A}_{i}},1\right)}{\Gamma\left(N_{s}\right)}.
\label{I2i_final}
\end{align}
Hence, taking into consideration \eqref{I1i_final}, \eqref{I2i_final} and since $U\left(m_{i}-1\right)U\left(m_{i}-2\right)=U\left(m_{i}-2\right)$, Eq. \eqref{Eq:test1} results to \eqref{Eq:F_Tk_final}. This concludes the proof.
\end{IEEEproof}

\begin{thm}
The CDF of the energy statistics assuming a non-ideal RF front-end and that the channel occupancy vector $\Theta=\tilde{\Theta}_{2,0}=\left[0, 0, 0, 0, 0, 0\right]$), can be obtained~by
\begin{align}
F_{T_k}\left(x\left|\tilde{\Theta}_{2,0}\right.\right) = \frac{\gamma\left(N_s,\frac{N_s x}{2\mathcal{A}_5}\right)}{\Gamma\left(N_s\right)}.
\label{Eq:CDF_non_ideal_all_idle}
\end{align}
\end{thm}
\begin{IEEEproof}
If the channel occupancy vector $\Theta$ is the all idle vector, i.e., $\Theta=\tilde{\Theta}_{2,0}=\left[0, 0, 0, 0, 0, 0\right]$, then, in accordance to \eqref{eq:sigma}, the signal variance can be expressed as
%\begin{align}
$\sigma^2_{\tilde{\Theta}_{2,0}} = \mathcal{A}_5.$
%\end{align}
According to \eqref{Eq:F_Tk_cond_H_Theta}, since $\sigma^2_{\tilde{\Theta}_{2,0}}$ is independent of $H$, the CDF of the energy statistics, assuming an non-ideal RF front-end, when all the channels of $\{-k-1, -k, -k+1, k-1, k, k+1\}$ are idle, can be obtained by  \eqref{Eq:CDF_non_ideal_all_idle}. This concludes the proof.
\end{IEEEproof}

Based on the above analysis, the detection probability of the energy detector
with RF impairments~is
\begin{equation}
\mathcal{P}_{D}=\sum_{i=1}^{\card\left(\tilde{\Theta}_{1}\right)}P_{r}\left(\tilde{\Theta}_{1}\right)\left(1-F_{T_{k}}\left(\gamma^{\text{ni}}\left|\tilde{\Theta}_{1}\right.\right)\right),\label{Eq:P_D_RF}
\end{equation}
where $\tilde{\Theta}_{1}$ is the set defined as
%\[
$\tilde{\Theta}_{1}=\left[\theta_{k}=1,\theta_{k-1},\theta_{k+1},\theta_{-k+1},\theta_{-k-1},\theta_{-k}\right].$
%\]
Similarly, the probability of false alarm is
\begin{align}
\mathcal{P}_{FA}=\sum_{i=1}^{\card\left(\tilde{\Theta}_{2,c}\right)} P_r\left(\tilde{\Theta}_{2}\right)\left(1-F_{T_{k}}\left(\gamma^{\text{ni}}\left|\tilde{\Theta}_{2,c}\right.\right)\right)
%\\
+ P_r\left(\tilde{\Theta}_{2,0}\right)\frac{\Gamma\left(N_{s},\frac{N_{s}x}{2\mathcal{A}_{5}}\right)}{\Gamma\left(N_{s}\right)},\label{Eq:P_F_RF}
\end{align}
where $\Pr\left(\Theta\right)$ denotes the probability of the given channel occupancy $\Theta$, $\tilde{\Theta}_{2,c}$ is the set defined~as
%\begin{equation}
$\tilde{\Theta}_{2,c}=\tilde{\Theta}_{2}-\tilde{\Theta}_{2,0},$
%\end{equation}
and $\tilde{\Theta}_{2}$ is the set defined as
%\begin{equation}
$\tilde{\Theta}_{2}=\left[\theta_{k}=0,\theta_{k-1},\theta_{k+1},\theta_{-k+1},\theta_{-k-1},\theta_{-k}\right].$
%\end{equation}
Note that \eqref{Eq:P_F_RF} applies even when the channel $K$ or $-K$ is sensed. However, in this case $\tilde{\Theta}_{1}$ and $\tilde{\Theta}_{2}$ can be obtained by
$\tilde{\Theta}_{1}=\left[\theta_{k}=1,\theta_{k-1},\theta_{k+1}=0,\theta_{-k+1},\theta_{-k-1}=0,\theta_{-k}\right]$ and $\tilde{\Theta}_{2}=\left[\theta_{k}=0, \theta_{k-1}, \theta_{k+1}=0, \theta_{-k+1}, \theta_{-k-1}=0, \theta_{-k}\right]$, respectively.
%whereas $P_r\left(\tilde{\Theta}=\left[\theta_{k},\theta_{k-1},\theta_{k+1}=1,\theta_{-k+1},\theta_{-k-1}=1,\theta_{-k}\right]\right)=0$.
%\subsection{Design curves}
%To derive the design curves, we observe that the expressions of the detection and false alarm probabilities are functions of the threshold $\gamma\left(k\right)$,
%\begin{align}
%\mathcal{P}_{D}&=\mathfrak{f}_{D}\left(\gamma\left(k\right)\right),
%\label{Eq:P_D_general_form}\\
%\mathcal{P}_{FA}&=\mathfrak{f}_{FA}\left(\gamma\left(k\right)\right),
%\label{Eq:P_FA_general_form}
%\end{align}
%where $\mathfrak{f}_{D}\left(\cdot\right)$ and $\mathfrak{f}_{FA}\left(\cdot\right)$ are given by (\ref{Eq:P_FA_Ideal_RF}) and (\ref{Eq:P_D_Ideal_RF}) in case of ideal RF front-end, and by (\ref{Eq:P_D_RF}) and (\ref{Eq:P_F_RF}) in case of non-ideal RF front-end. The threshold $\gamma\left(k\right)$ can be expressed, using the inverse form of (\ref{Eq:P_D_general_form}) and (\ref{Eq:P_FA_general_form}), as
%\begin{align}
%\mathfrak{f}_{D}^{-1}\left(\mathcal{P}_{D}\right) = \mathfrak{f}_{FA}^{-1}\left(\mathcal{P}_{FA}\right).
%\end{align}

\section{Cooperative Spectrum Sensing with Decision Fusion}\label{sec:Cooperative_Spectrum_Sensing}

In this section, we consider a cooperative spectrum sensing scheme, in which each SU makes a binary decision on the channel occupancy, namely `0' or `1' for the absence or presence of PU activity, respectively, and the one-bit individual decisions are forwarded to a  FC over a narrowband reporting channel. The sensing channels (the channels between the PU and the SUs) are considered identical and independent. Moreover, we assume that the decision device of the FC is implemented with the $k_{\rm{SU}}$-out-of-$n_{\rm{SU}}$ rule, which implies that if there are $k_{\rm{SU}}$ or more SUs that individually decide that the channel is busy, the FC decides that the channel is occupied. Note that when $k_{\rm{su}}=1$, $k_{\rm{su}}=n_{\rm{su}}$ or $k_{\rm{su}}=\lceil n/2\rceil$, the $k_{\rm{su}}$-out-of-$n_{\rm{su}}$ rule is simplified to the OR rule, AND rule and Majority rule, respectively.

\subsection{Ideal RF Front-End}

Here, we derive closed form expression for the false alarm and detection probabilities, assuming that the RF front-ends of the SUs are ideal, considering both scenarios of error free and imperfect reporting channels.

\subsubsection{Reporting Channels without Errors}
If the channel between the SUs and the FC is error free, the false alarm probability (${\cal P}_{C,fa}$) and the detection probability (${\cal P}_{C,d}$) are given  by \cite[Eq. (17)]{ED_Cooperative_Spectrum_Sensing_in_CR}
\begin{align}
{\cal P}_{C,fa}\hspace{-0.15cm}=\hspace{-0.15cm} \sum_{i=k_{\rm{su}}}^{n_{\rm{su}}}\left(\begin{array}{c}n_{\rm{su}}\\i\end{array} \right) \left({\cal P}_{fa}\right)^{i} \left(1-{\cal P}_{fa}\right)^{n_{\rm{su}}-i}
\text{ and }
{\cal P}_{C,d}\hspace{-0.15cm}=\hspace{-0.15cm} \sum_{i=k_{\rm{su}}}^{n_{\rm{su}}}\left(\begin{array}{c}n_{\rm{su}}\\i\end{array} \right) \left({\cal P}_{d}\right)^{i} \left(1-{\cal P}_{d}\right)^{n_{\rm{su}}-i}.\label{Eq:P_FA_C_D_ideal}
\end{align}
Taking into consideration (\ref{Eq:P_FA_Ideal_RF}) (\ref{Eq:P_D_Ideal_RF}) and (\ref{CDF_theta_1_ideal}) and after some basic algebraic manipluations, Eqs.~(\ref{Eq:P_FA_C_D_ideal}) can be expressed as
\begin{align}
{\cal P}_{C,fa}&= \sum_{i=k_{\rm{su}}}^{n_{\rm{su}}}\left(\begin{array}{c}n_{\rm{su}}\\i\end{array} \right)\left(\frac{\Gamma\left(N_{s},\frac{N_{s}\gamma\left(k\right)}{\sigma_{w}^{2}}\right)}{\Gamma\left(N_{s}\right)}\right)^{i}
%\nonumber\\&
\left(\frac{\gamma\left(N_{s},\frac{N_{s}\gamma\left(k\right)}{\sigma_{w}^{2}}\right)}{\Gamma\left(N_{s}\right)}\right)^{n-i},
\\
{\cal P}_{C,d}=
&\sum_{i=k_{\rm{su}}}^{n_{\rm{su}}}\left(\begin{array}{c}n_{\rm{su}}\\i\end{array} \right)
\left({\exp\left(\frac{\sigma_{w}^2}{\sigma_h^2 \sigma_s^2}\right)} \sum_{k=0}^{N_s-1}\frac{1}{k!} \left(\frac{N_s \gamma}{\sigma_h^2 \sigma_s^2}\right)^k \Gamma\left(-k+1,\frac{\sigma_w^2}{\sigma_h^2 \sigma_s^2},\frac{N_s \gamma}{\sigma_h^2 \sigma_s^2},1\right)\right)^{i}
\nonumber\\& \times
\left(1-{\exp\left(\frac{\sigma_{w}^2}{\sigma_h^2 \sigma_s^2}\right)} \sum_{k=0}^{N_s-1}\frac{1}{k!} \left(\frac{N_s \gamma}{\sigma_h^2 \sigma_s^2}\right)^k \Gamma\left(-k+1,\frac{\sigma_w^2}{\sigma_h^2 \sigma_s^2},\frac{N_s \gamma}{\sigma_h^2 \sigma_s^2},1\right)\right)^{n_{\rm{su}}-i}.
\end{align}

\subsubsection{Reporting Channels with Errors}

If the reporting channel is imperfect, error occur on the detection of the transmitted, by the SU, bits. In this case, the false alarm and the detection probabilities can be derived by \cite[Eq. (18)]{ED_Cooperative_Spectrum_Sensing_in_CR}
\begin{align}
{\cal P}_{C,{\cal X}} = \sum_{i=k_{\rm{su}}}^{n_{\rm{su}}} \left(\begin{array}{c}n\\i \end{array}\right) \left({\cal P}_{{\cal X},e}\right)^{i} \left(1-{\cal P}_{{\cal X},e}\right)^{n_{\rm{su}}-i},
\label{Eq:PCX}
\end{align}
where
%\begin{align}
${\cal P}_{{\cal X},e} = {\cal P}_{{\cal X}} \left(1-P_{e}\right) + \left(1-{\cal P}_{{\cal X}}\right) P_{e},$
%\label{Eq:P_C_X_with_errors}
%\end{align}
is the equivalent false alarm (`${\cal X}=fa$') or detection (`${\cal X}=d$') probability and $P_{e}$ is the cross-over probability of the reporting channel, which is equal to the bit error rate (BER) of the channel. Considering binary phase shift keying (BPSK), ideal RF front-end in the FC and Rayleigh fading, the BER can be expressed as
\begin{align}
P_{e} = \frac{1}{2}\left(1-\sqrt{\frac{\gamma_{r}}{1+\gamma_{r}}}\right),
\end{align}
with $\gamma_{r}$ be the signal to noise ratio (SNR) of the link between the SUs and the FC.

%Notice that since ${\cal P}_{{\cal X}}\in\left[0,1\right]$, ${\cal P}_{{\cal X},e}$ is bounded in $\left[P_{e},1-P_{e}\right]$. Consequently, according to \eqref{Eq:PCX}, ${\cal P}_{C,{\cal X}}\in\left[{\cal P}_{C,{\cal X}}^{-},{\cal P}_{C,{\cal X}}^{+}\right]$, where
%\begin{align}
%{\cal P}_{C,{\cal X}}^{-}\hspace{-0.1cm}=\hspace{-0.1cm}\sum_{i=k_{\rm{su}}}^{n_{\rm{su}}} \left(\begin{array}{c}n_{\rm{su}}\\i \end{array}\right) \left(
%P_{e}\right)^{i} \left(1-P_{e}\right)^{n_{\rm{su}}-i}
%\text{ and }
%{\cal P}_{C,{\cal X}}^{+}\hspace{-0.1cm}=\hspace{-0.1cm}\sum_{i=k_{\rm{su}}}^{n_{\rm{su}}} \left(\begin{array}{c}n_{\rm{su}}\\i \end{array}\right) \left(1-P_{e}\right)^{i} \left(P_{e}\right)^{n_{\rm{su}}-i}.
%\label{Eq:P_C_X_bounds_ideal}
%\end{align}

\subsection{Non-Ideal RF Front-End}

In this subsection, we consider that the RXs front-end of the SUs suffer from different level RF imperfections.

\subsubsection{Reporting Channels without Errors}
In this section, we assume that the reporting channel is error free and
%In case of an error free reporting channel, and that
%the false alarm probability and the detection probability of the SU $i$ are ${\cal P}_{fa, i}$ and ${\cal P}_{d, i}$, respectively. We consider
that the SU $j$ sends $d_{j,k}=0$ or $d_{j,k}=1$ to the FC to report absence or presence of PU activity at the channel $k$.

If the sensing channel $k$ is idle ($\theta_{k}=0$), then the probability that the $j^{\text{th}}$ SU reports that the channel is busy ($d_{j,k} =1$), can be expressed as ${\cal P}_{fa, j}$, while the probability that the $j^{\text{th}}$ SU reports that the channel is idle ($d_{j,k} =0$), is given by $\left(1-{\cal P}_{fa, j}\right)$. Therefore, since each SU decides individually whether there is PU activity in the channel $k$, the probability that the $n$ SUs report a given decision set $\mathcal{D}=\left[d_{1,k}, d_{2,k}, \cdots, d_{n_{\rm{su}},k}\right]$, if $\theta_{k}=0$, can be written~as
\begin{align}
{\cal{P}}_{fa}(\mathcal{D})=\prod_{j=1}^{n_{\rm{su}}}
\left(
U\left(-d_{j,k}\right)\left(1-{\cal P}_{fa, j}\right)
+U\left(d_{j,k}-1\right){\cal P}_{fa, j}\right).
\end{align}
Furthermore, based on the \rm{$k_{\rm{su}}$-out-of-$n_{\rm{su}}$} rule, the FC decides that the $k^{\text{th}}$ channel is busy, if the $k_{\rm{su}}$ out of the $n_{\rm{su}}$ SUs reports ``1''. Consequently, for a given decision set, the false alarm probability at the FC can be evaluated by
\begin{align}
{\cal P}_{C,FA\left|\mathcal{D}\right.} =
U\left(\sum_{l=1}^{n_{\rm{su}}}d_{l,k}-k_{\rm{su}}\right)
\prod_{j=1}^{n_{\rm{su}}}
\left(
U\left(- d_{j,k}\right)
\left(1-{\cal P}_{fa, j}\right)
+U\left(d_{j,k}-1\right)
{\cal P}_{fa, j}
\right).
\end{align}
Hence, for any possible $\mathcal{D}$, the false alarm at the FC, using \rm{$k_{\rm{su}}$-out-of-$n_{\rm{su}}$} rule, can be obtained~by
\begin{align}
{\cal P}_{C,FA} = \sum_{i=1}^{\card\left(\mathcal{D}\right)}
U\left(\sum_{l=1}^{n_{\rm{su}}}d_{l,k}-k_{\rm{su}}\right)
\prod_{j=1}^{n_{\rm{su}}}
\left(
U\left(-d_{j,k}\right)\left(1-{\cal P}_{fa, j}\right)
+U\left(d_{j,k}-1\right){\cal P}_{fa, j}\right).
\label{Eq:Cooperative_FA_dif_imp_general_no_error}
\end{align}

Similarly, the detection probability at the FC, using $k_{\rm{su}}$-out-of-$n_{\rm{su}}$ rule, can be expressed as
\begin{align}
{\cal P}_{C,D} = \sum_{i=1}^{\card\left(\mathcal{D}\right)}
U\left(\sum_{l=1}^{n_{\rm{su}}}d_{l,k}-k_{\rm{su}}\right)
\prod_{j=1}^{n_{\rm{su}}}
\left(
U\left(-d_{j,k}\right)\left(1-{\cal P}_{d, j}\right)
+U\left(d_{j,k}-1\right){\cal P}_{d, j}\right).
\label{Eq:Cooperative_D_dif_imp_general_no_error}
\end{align}

Note that if the FC uses the OR rule, Eqs. \eqref{Eq:Cooperative_FA_dif_imp_general_no_error} and \eqref{Eq:Cooperative_D_dif_imp_general_no_error} can be simplified~to
\begin{align}
{\cal P}_{\text{OR},FA}=1-\prod_{i=1}^{n_{\rm{su}}} \left(1-{\cal P}_{fa, i}\right),
\text{ and }
{\cal P}_{\text{OR},D}=1-\prod_{i=1}^{n_{\rm{su}}} \left(1-{\cal P}_{d, i}\right),
\end{align}
respectively, while  if the FC uses the AND rule, Eqs. \eqref{Eq:Cooperative_FA_dif_imp_general_no_error} and \eqref{Eq:Cooperative_D_dif_imp_general_no_error} can be simplified~to
\begin{align}
{\cal P}_{\text{AND},FA}=\prod_{i=1}^{n_{\rm{su}}} {\cal P}_{fa, i},
\text{ and }
{\cal P}_{\text{AND},D}=\prod_{i=1}^{n_{\rm{su}}} {\cal P}_{d, i},
\end{align}
respectively.

In the special case where all the SUs suffer from the same level of RF impairments, the false alarm probability (${\cal P}_{C,fa}$) and the detection probability (${\cal P}_{C,d}$) are given by
\begin{align}
{\cal P}_{C,FA} \hspace{-0.1cm}=\hspace{-0.1cm} \sum_{i=k_{\rm{su}}}^{n_{\rm{su}}}\left(\begin{array}{c}n_{\rm{su}}\\i\end{array} \right) \left({\cal P}_{FA}\right)^{i} \left(1-{\cal P}_{FA}\right)^{n_{\rm{su}}-i},\label{Eq:P_C_FA_RF}
\text{ and }
{\cal P}_{C,D} \hspace{-0.1cm}=\hspace{-0.1cm} \sum_{i=k_{\rm{su}}}^{n_{\rm{su}}}\left(\begin{array}{c}n_{\rm{su}}\\i\end{array} \right) \left({\cal P}_{D}\right)^{i} \left(1-{\cal P}_{D}\right)^{n_{\rm{su}}-i},
%\label{Eq:P_C_D_RF}
\end{align}
where ${\cal P}_{FA}$ and ${\cal P}_{D}$ are given by (\ref{Eq:P_F_RF}) and (\ref{Eq:P_D_RF}), respectively.

\subsubsection{Reporting Channels with Errors}

Next, we consider and imperfect reporting channel. In this scenario, the false alarm and the detection probabilities can be derived by
\begin{align}
{\cal P}_{C,\mathcal{X}} = \sum_{i=1}^{\card\left(\mathcal{D}\right)}
U\left(\sum_{l=1}^{n_{\rm{su}}}d_{l,k}-k_{\rm{su}}\right)
\prod_{j=1}^{n_{\rm{su}}}
\left(
U\left(-d_{j,k}\right)\left(1-{\cal P}_{\mathcal{X},e, j}\right)
+U\left(d_{j,k}-1\right){\cal P}_{\mathcal{X},e, j}\right),
\label{Eq:Cooperative_dif_imp_general_error}
\end{align}
where ${\cal P}_{{\cal X},e,j}$ can be derived by
\begin{align}
{\cal P}_{{\cal X},e,j} = {\cal P}_{{\cal X},j} \left(1-P_{e,j}\right) + \left(1-{\cal P}_{{\cal X},j}\right) P_{e,j},
\label{PXej}
\end{align} with ${\cal P}_{{\cal X},j}$ denoting the equivalent false alarm (`${\cal X}=FA$') or detection (`${\cal X}=D$') probability of the $j^{\text{th}}$ SU and $P_{e,j}$ being the cross-over probability of the reporting channel connecting the $j^{\text{th}}$ SU with the FC. Notice that since ${\cal P}_{{\cal X},j}\in\left[0,1\right]$, based on \eqref{PXej} ${\cal P}_{{\cal X},e,j}$ is bound by $P_{e,j}$ and $1-P_{e,j}$.

In the special case where all the SUs suffer from the same level of RF impairments, Eq. \eqref{Eq:Cooperative_dif_imp_general_error} can be expressed as \cite[Eq. (18)]{ED_Cooperative_Spectrum_Sensing_in_CR}
\begin{align}
{\cal P}_{C,{\cal X}} = \sum_{i=k_{\rm{su}}}^{n_{\rm{su}}} \left(\begin{array}{c}n_{\rm{su}}\\i \end{array}\right) \left({\cal P}_{{\cal X},e}\right)^{i} \left(1-{\cal P}_{{\cal X},e}\right)^{n_{\rm{su}}-i}.
\label{PcX}
\end{align}
%Moreover, according to \eqref{PcX}, since ${\cal P}_{{\cal X},e,j}$ is bounded in $\left[P_{e,j}, 1-P_{e,j}\right]$, ${\cal P}_{C,{\cal X}}$ is bounded by
%\begin{align}
%{\cal P}_{C,{\cal X}}^{-}\hspace{-0.1cm}=\hspace{-0.1cm}\sum_{i=k_{\rm{su}}}^{n_{\rm{su}}} \left(\begin{array}{c}n_{\rm{su}}\\i \end{array}\right) \left(
%P_{e}\right)^{i} \left(1-P_{e}\right)^{n_{\rm{su}}-i}
%\text{ and }
%{\cal P}_{C,{\cal X}}^{+}\hspace{-0.1cm}=\hspace{-0.1cm}\sum_{i=k_{\rm{su}}}^{n_{\rm{su}}} \left(\begin{array}{c}n_{\rm{su}}\\i \end{array}\right) \left(1-P_{e}\right)^{i} \left(P_{e}\right)^{n_{\rm{su}}-i}.
%\label{Eq:P_C_X_bounds}
%\end{align}
%Eqs. \eqref{Eq:P_C_X_bounds} indicate that even if the SUs can be considered ideal, the spectrum sensing capabilities of the FC degrade due to the imperfect reporting channels, especially in case of a high cross-over probability, i.e. low $\rm{SNR}$ of the reporting channels.

\section{Numerical and Simulation Results}\label{sec:Numerical_Results}

In this section, we investigate the effects of RF impairments on the spectrum sensing performance of EDs by illustrating analytical and Monte-Carlo simulation results for different RF imperfection levels. In particular, we consider the following insightful scenario. It is assumed that there are $K=8$ channels and the second channel is sensed (i.e., $k=2$). The signal and the total guard band bandwidths are assumed to be $W_{sb}=1\text{ }\rm{MHz}$ and $W_{gb}=125\text{ }\rm{KHz}$, respectively, while the sampling rate is chosen to be equal to the bandwidth of wireless signal as $W=9\text{ }\rm{MHz}$. Moreover, the channel occupancy process is assumed to be Bernoulli distributed with probability, $q=1/2$, and independent across channels, while the signal variance is equal for all channels. The number of samples is set to $5$ ($N_s=5$), while it is assumed that $\sigma_{h}^{2}=\sigma_{w}^{2}=1$.
In addition, for simplicity and without loss of generality, we consider an ideal clipping PA.
In the following figures, the numerical results are shown with continuous lines, while markers are employed to illustrate the simulation results.
Moreover, the performance of a classical ED with ideal RF front-end is used as a~benchmark.

\begin{figure}
\centering\includegraphics[width=0.75\linewidth,trim=0 0 0 0,clip=false]{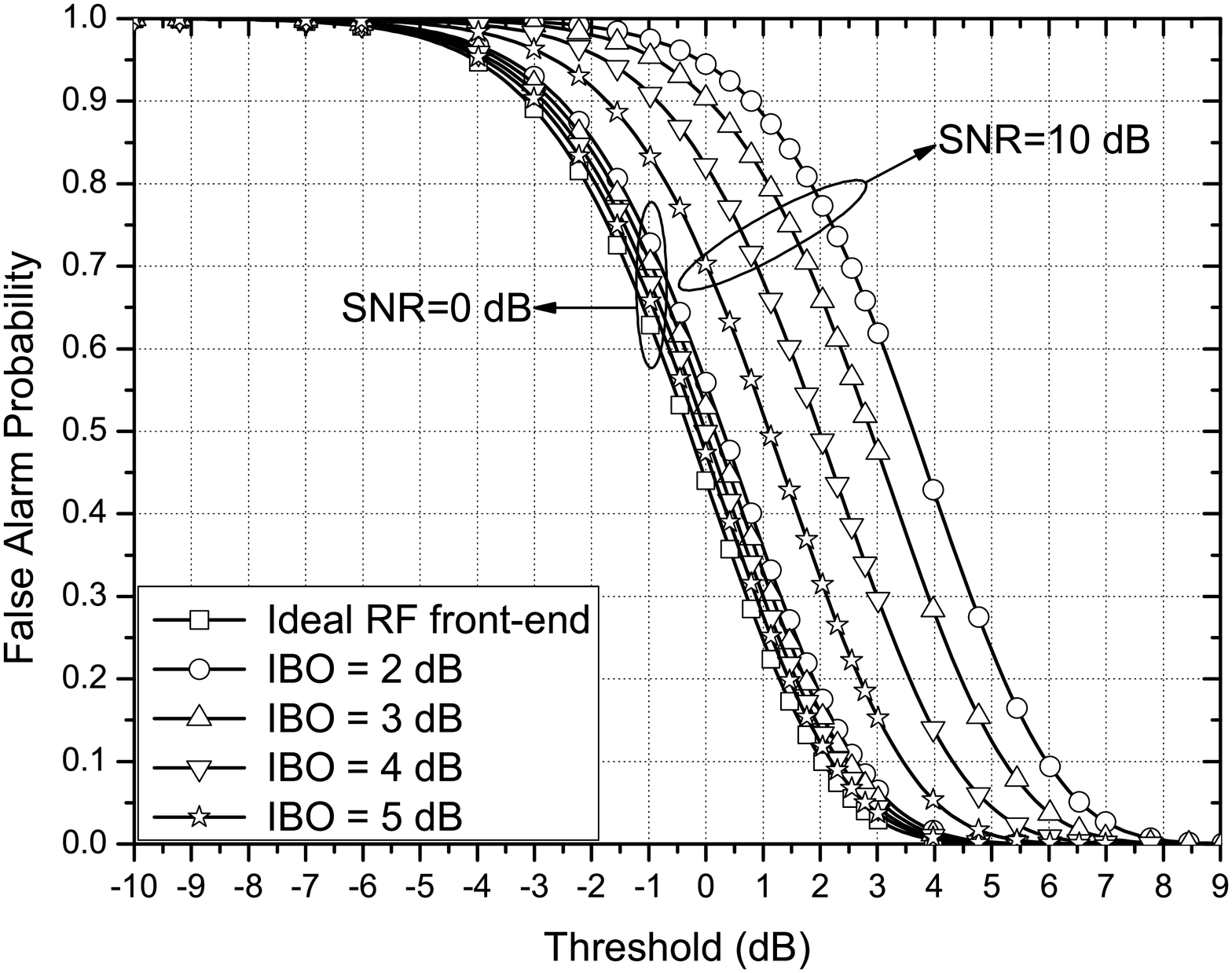}
\vspace{-0.43cm}
\caption{False alarm probability vs Threshold for different values of IBO and occupied channel $\rm{SNR}$ values, when $\rm{IRR}$ and $\beta$ are considered to be equal to $25\rm{dB}$ and $100\rm{Hz}$, respectively.}
\label{fig:FA_Thershold_IBO_SNR}
\vspace{-0.8cm}
\end{figure}

\begin{figure}
\centering\includegraphics[width=0.75\linewidth,trim=0 0 0 0,clip=false]{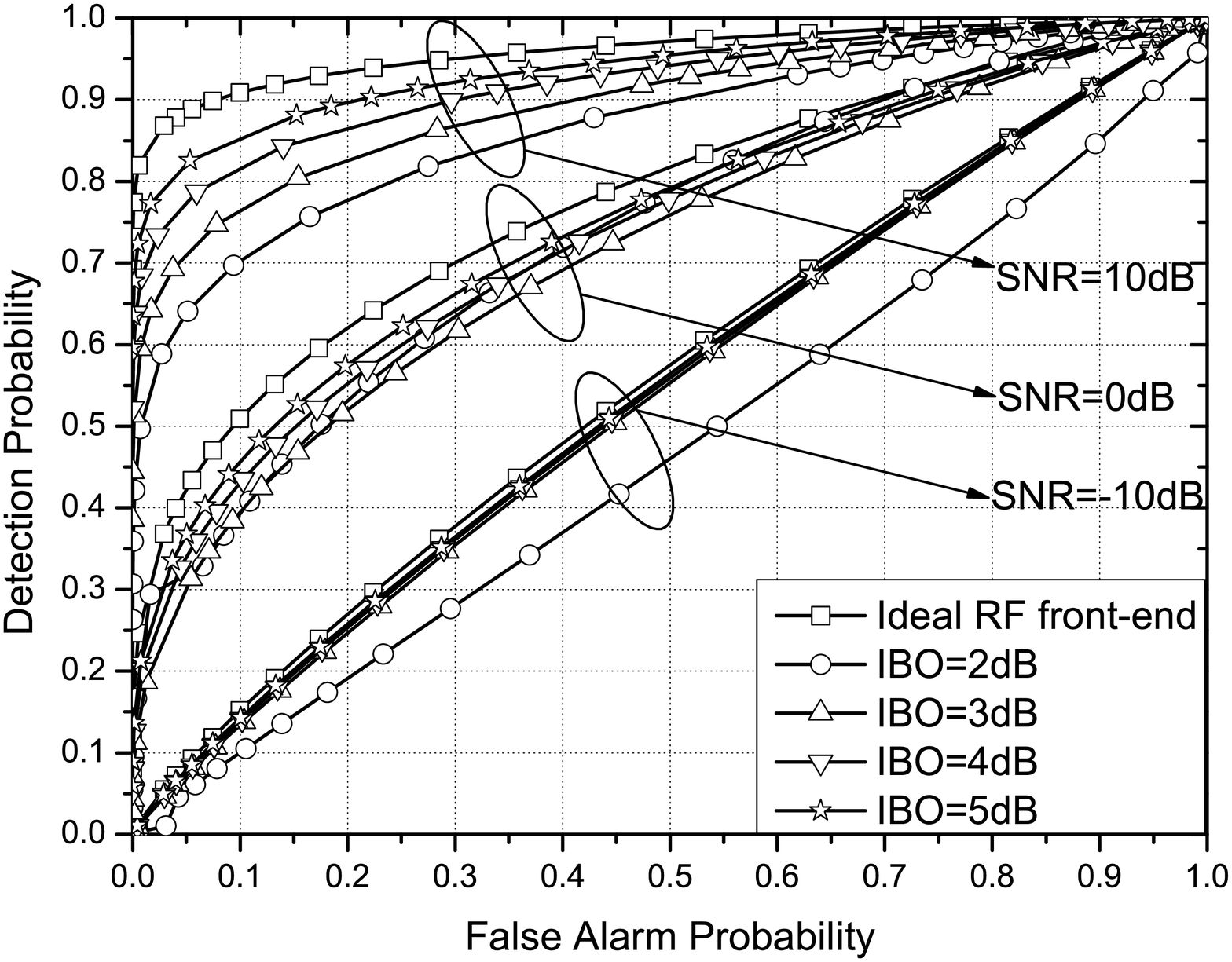}
\vspace{-0.43cm}
\caption{ROC for different values of IBO and occupied channel SNR values, when $IRR$ and $\beta$ are considered to be equal to $25\rm{dB}$ and $100\rm{Hz}$, respectively.}
\label{fig:ROC_IBO_SNR}
\vspace{-0.8cm}
\end{figure}

Figs. \ref{fig:FA_Thershold_IBO_SNR} and \ref{fig:ROC_IBO_SNR} demonstrate the impact of LNA non-linearities on the performance of the classical ED, assuming different $\rm{SNR}$ values.
Specifically, in Fig. \ref{fig:FA_Thershold_IBO_SNR}, false alarm probabilities are plotted against threshold for different $\rm{SNR}$ and $\rm{IBO}$ values, considering $\beta = 100\text{ }\rm{Hz}$, $\rm{IRR}=25\text{ }\rm{dB}$ and phase imbalance equal to $\phi=3^{o}$.
It becomes evident from this figure that the analytical results are identical with simulation results; thus, verifying the presented analytical framework.
Additionally, it is observed that for a given $\rm{IBO}$ value, as $\rm{SNR}$ increases, the interference for the neighbor and mirror channels increases; hence, the false alarm probability increases. On the contrary as $\rm{IBO}$ increases, for a given $\rm{SNR}$ value, the effects of LNA non-linearities are constrained, and therefore the false alarm probability decreases.

In Fig. \ref{fig:ROC_IBO_SNR}, receiver operation curves (ROCs) are plotted for different $\rm{SNR}$ and $\rm{IBO}$ values, considering the $\beta=100\text{ }\rm{Hz}$, $\rm{IRR}=25\text{ }\rm{dB}$ and $\phi=3^o$.
We observe that for low $\rm{SNR}$ values, LNA non-linearities do not affect the ED performance. However, as $\rm{SNR}$ increases, the distortion noise caused due to the imperfection of the amplifier increases; as a result, LNA non-linearities become to have more adverse effects on the spectrum capabilities of the classical ED, significantly reducing its performance for low $\rm{IBO}$ values. Furthermore, as $\rm{IBO}$ increases, the effects of LNA non-linearities become constrained and therefore the performance of the non-ideal ED tends to the performance of the ideal~ED.

\begin{figure}
\centering\includegraphics[width=0.75\linewidth,trim=0 0 0 0,clip=false]{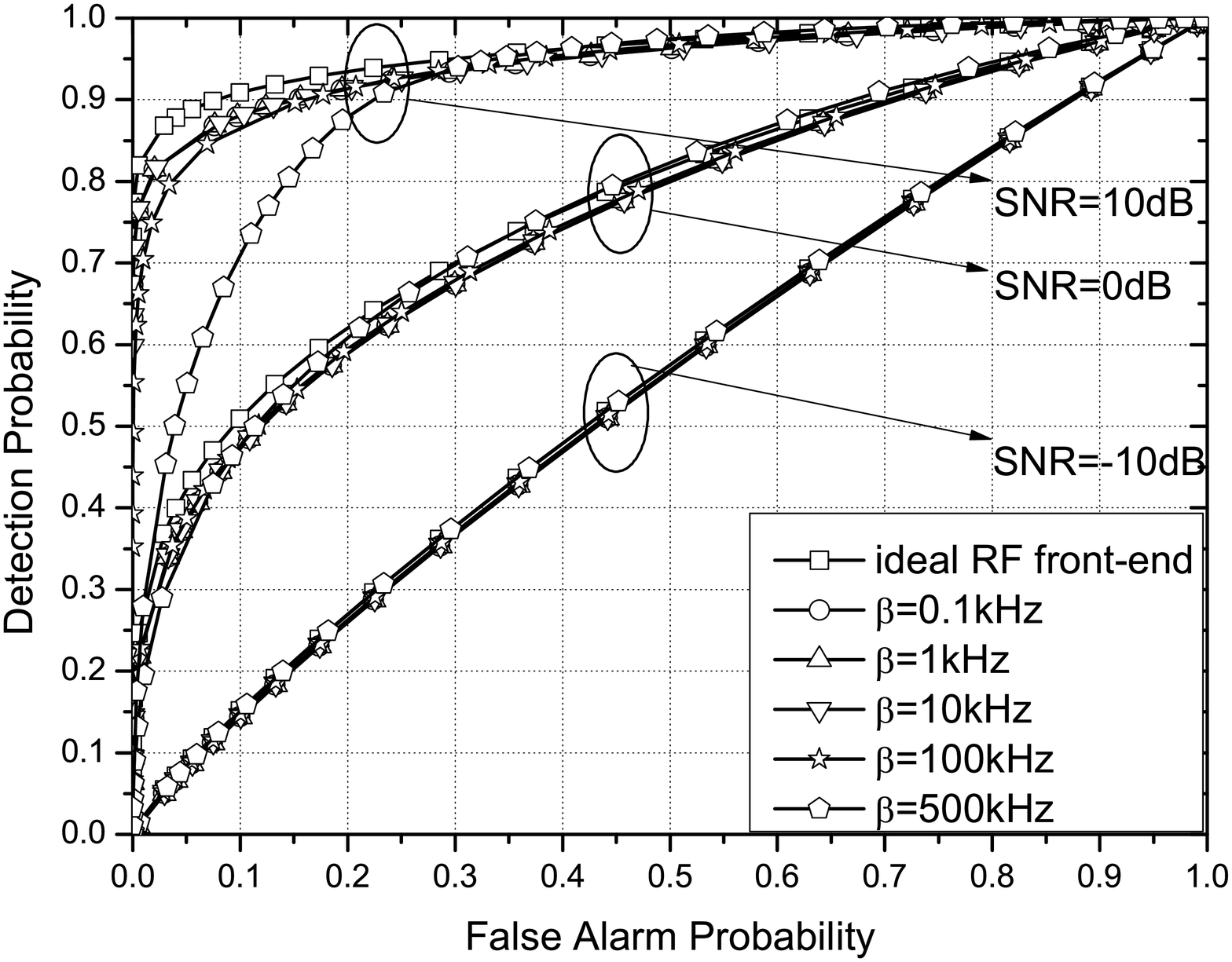}
\vspace{-0.43cm}
\caption{ROCs for different values of $\beta$ and occupied channel SNR values, when $\IBO$ and $\rm{IRR}$ are considered to be equal to $6\rm{dB}$ and $25\rm{dB}$, respectively.}
\label{fig:ROC_beta_IBO_6dB}
\vspace{-0.8cm}
\end{figure}
Fig.\ref{fig:ROC_beta_IBO_6dB} illustrates the impact of PHN  on the performance of the classical ED, assuming various $\rm{SNR}$ values, when $\rm{IRR} = 25\text{ }\rm{dB}$,  $\phi=3^{o}$ and $\rm{IBO}=6\text{ }\rm{dB}$. We observe that for practical levels of IQI and PHN, the signal leakage from channels $-k+1$ and $-k-1$ to channel $-k$ due to PHN is small, therefore the signal leakage to channel $k$ from the channel $-k-1$ and $-k+1$ due to the joint effect of PHN and IQI is in the range of $\left[-70\text{ }\rm{dB},-50\text{ }\rm{dB} \right]$. Consequently, in the low $\rm{SNR}$ regime the leakage from the channels $-k-1$ and $-k+1$ do not affect the spectrum sensing capabilities.
Hence, it becomes evident that at low $\rm{SNR}$ values, PHN do not affect the spectrum sensing capability of the classical ED compared with the ideal RF front-end ED.
On the other hand, as $\rm{SNR}$ increases, PHN has more detrimental effects on the spectrum sensing capabilities of the classical ED, significantly reducing the ED performance for high $\beta$ values.
%Furhtermore, in the low false alarm probability regime, for a given $\rm{SNR}$, as $\beta$ increases, the interference from the channels $-k-1$, $-k+1$, $k-1$ and $k+1$ increases, and consequently the effects of PHN become more severe.

\begin{figure}
\centering\includegraphics[width=0.75\linewidth,trim=0 0 0 0,clip=false]{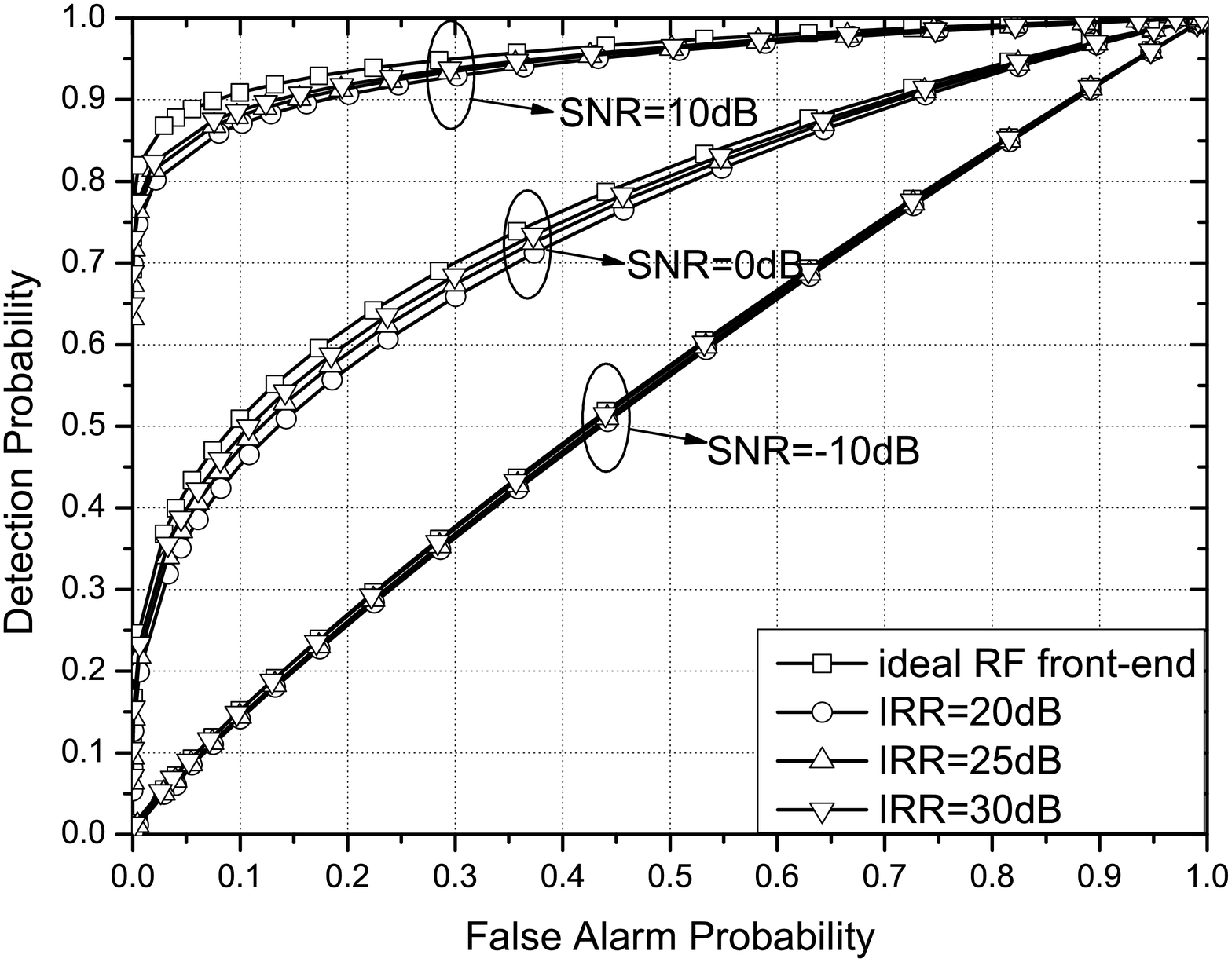}
\vspace{-0.43cm}
\caption{ROCs for different values of $IRR$ and occupied channel SNR values, when $\IBO$ and $\beta$ are considered to be equal to $6\rm{dB}$ and $100\rm{Hz}$, respectively.}
\label{fig:ROC_vs_IRR}
\vspace{-0.8cm}
\end{figure}
The effects of IQI on the spectrum sensing performance of ED are presented at Fig. \ref{fig:ROC_vs_IRR}. In particular, in this figure, ROCs are plotted assuming various $\rm{SNR}$s, when the $IBO=6\text{ }\rm{dB}$ and $\beta = 100Hz$.
Again, the analytical results coincide with simulation results, verifying the derived expressions. Moreover, at low $\rm{SNR}$s, it is observed that there is no significant performance degradation due to IQI. Nonetheless, as $\rm{SNR}$ increases, the interference of the mirror channels increases and as a result this RF imperfection notably affects the spectrum sensing performance.
Additionally, for a given $\rm{SNR}$, we observe that as $\rm{IRR}$ increases, the signal leakage of the mirror channels, due to IQI, decreases; hence, the performance of the non-ideal ED tends to become identical to the one of the ideal ED.
Finally, when compared with the spectrum sensing performance affected by
LNA nonlinearities, as depicted in Fig. \ref{fig:ROC_IBO_SNR}, it becomes apparent that the impact of LNA nonlinearity to the spectrum sensing performance is more detrimental than the impact of~IQI.

\begin{figure}
\centering\includegraphics[width=0.75\linewidth,trim=0 0 0 0,clip=false]{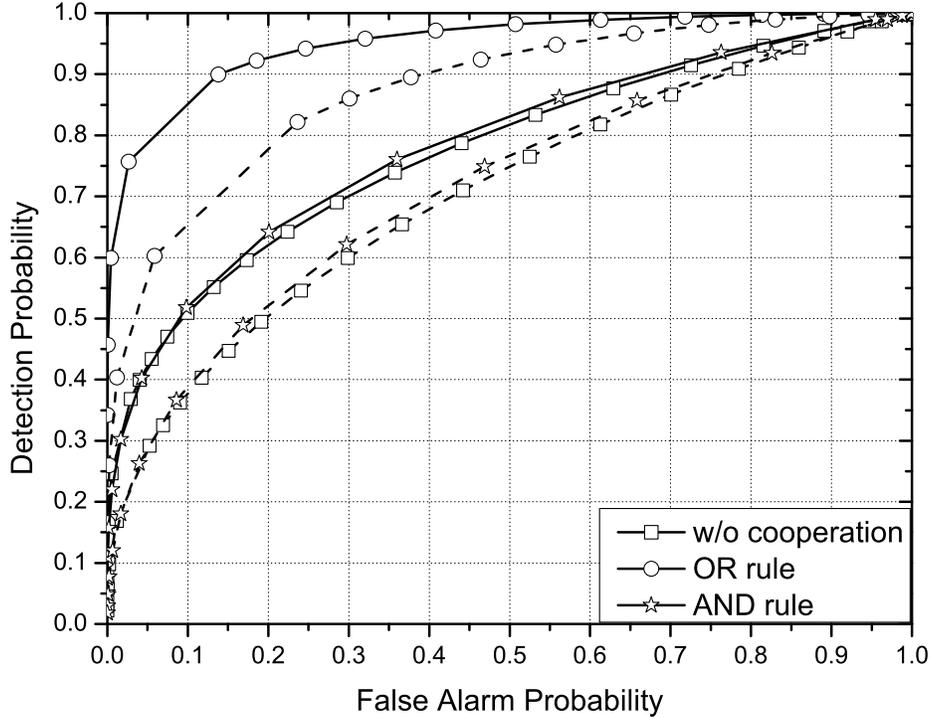}
\vspace{-0.43cm}
\caption{ROCs for ideal (continuous line) and non-ideal (dashed lines) RF front-end , when the CR network is equipped with $5$ SUs, $\rm{SNR}=0\text{ }\rm{dB}$, the reporting channel is considered error free, and $IBO=3\text{ }\rm{dB}$, $IRR=20\text{ }\rm{dB}$, and $\beta=100\text{ }\rm{Hz}$, for all the~SUs.}
\label{fig:ROC_cooperative_error_free_n5}
\vspace{-0.8cm}
\end{figure}
The effects of RF impairments in cooperative sensing, when the reporting channel is considered error free, is illustrated in Fig. \ref{fig:ROC_cooperative_error_free_n5}. In this figure, ROCs for ideal (continuous lines) and non-ideal (dashed lines) RF front-end SUs are presented, considering a CR network composed of $n_{\rm{su}}=5$ SUs, and a FC, which uses the OR or AND rule to decide whether the sensing channel is idle or busy. The EDs of the SUs are assumed identical with $\rm{IBO}=3\text{ }\rm{dB}$, $\rm{IRR}=20\text{ }\rm{dB}$, and $100\text{ }\rm{Hz}$ $3\text{ }\rm{dB}$ bandwidth. Again it is shown that the analytical results are identical with simulation results; thus, verifying the presented analytical framework.
%Moreover, it is observed that as $k_{\rm{SU}}$ decreases, the FC decision rule become more strict, and therefore the performance of the CR network improves.
When a given decision rule is applied, it becomes evident from the figure that the RF imperfections cause severe degradation of the sensing capabilities of the CR network. For instance, if the OR rule is employed and false alarm probability is equal to $14\%$, the RF impairments results to about $31\%$ degradation compared with the ideal RF front-end~scenario. This result indicates that it is important to take into consideration the hardware constraints of the low-cost spectrum sensing~SUs.

\begin{figure}
\centering\includegraphics[width=0.75\linewidth,trim=0 0 0 0,clip=false]{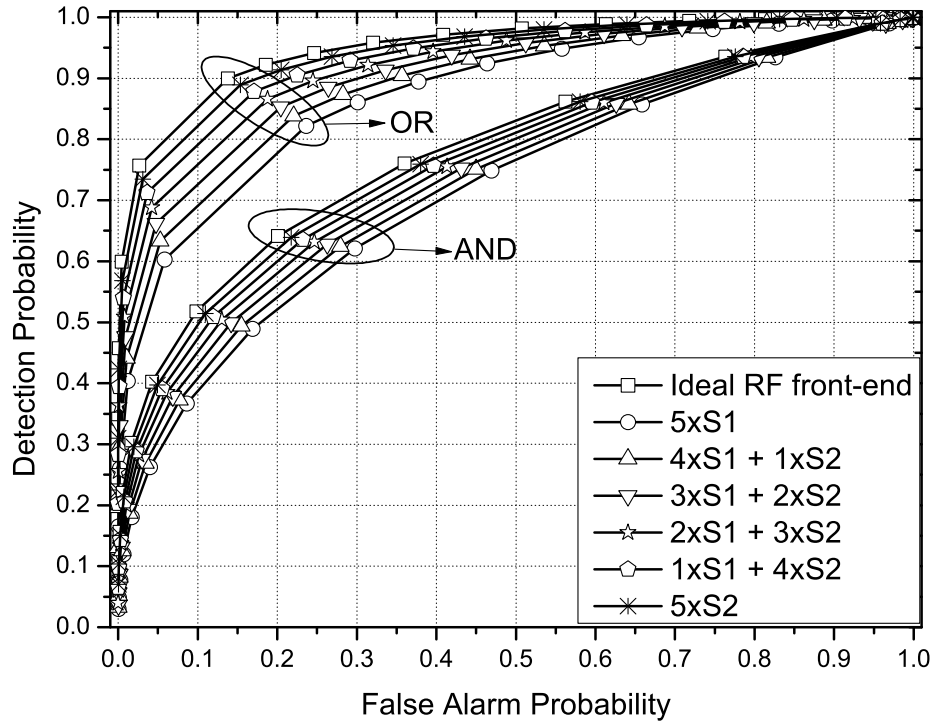}
\vspace{-0.43cm}
\caption{ROCs for ideal and non-ideal RF front-end, when the CR network is equipped with $5$ SUs under different levels of RF imperfections, $\rm{SNR}=0\text{ }\rm{dB}$, the reporting channel is considered error free and the FC uses $\rm{AND}$ or $\rm{OR}$ rule. $\text{S}_{1}$ and $\text{S}_{2}$ stands for SUs with $\rm{IBO}=3\text{ }\rm{dB}$ and $\rm{IRR}=20\text{ }\rm{dB}$, and $\rm{IBO}=6\text{ }\rm{dB}$ and $\rm{IRR}=300\text{ }\rm{dB}$, respectively.}
\label{fig:ROC_cooperative_error_free_n5_diff_RF}
\vspace{-0.8cm}
\end{figure}
In Fig. \ref{fig:ROC_cooperative_error_free_n5_diff_RF}, ROCs are illustrated for a CR network composed of $n=5$ SUs, which suffer from different levels of RF imperfections, and a FC that employs the $\rm{AND}$ or the $\rm{OR}$ rule to decide whether the sensing channel is idle or busy. In this scenario, we consider two types of SUs, namely $S_1$ and $S_2$. The RF front-end specifications of $S_1$ are $\rm{IBO}=3\text{ }\rm{dB}$, $\rm{IRR} = 20\text{ }\rm{dB}$ and $\beta = 100\text{ }\rm{Hz}$, whereas the specifications of $S_2$ are $\rm{IBO}=6\text{ }\rm{dB}$, $\rm{IRR} = 30\text{ }\rm{dB}$ and $\beta = 100\text{ }\rm{Hz}$. In other words, the CR network, in this scenario, includes both SUs of almost the worst ($S_1$) and almost optimal ($S_2$) quality. As benchmarks, the ROCs of a CR network equipped with classical ED sensor nodes in which the RF front-end is considered to be ideal, and CR networks that uses only $S_1$ or only $S_2$ sensor nodes are presented. In this figure, we observe the detrimental effects of the RF imperfections of the ED sensor nodes to the sensing capabilities of the CR network.
Furthermore, it is demonstrated that as the numbers of $S_1$ and $S_2$ SUs are decreasing and increasing respectively, the energy detection performance of the FC tends to become identical to the case when all the SUs are considered to be ideal.
This was expected since $S_2$ SUs have higher quality RF front-end characteristics than the other set of~SUs.

%\subsection{Discussion}

%Notice that for practical levels of IQI and PHN, the signal leakage from channels $-k+1$ and $-k+1$ to channel $-k$ due to PHN is small (for instance in order of $-30$ $\rm{dB}$), therefore the signal leakage to channel $k$ from the channel $-k+1$ and $-k-1$ due to the joint effect of PHN and IQI is in the range of $\left[-70\text{ }\rm{dB},-50\text{ }\rm{dB} \right]$.
%Hence, the term $\left|K_{2}\right|^{2}\sigma_{\psi\left|H_{-k},\Theta_{-k}\right.}^{2}$ should only effect if the channel coefficient and/or the signal strength at channels  $-k+1$ and/or $-k-1$ are extremely strong.
%Therefore, the terms referred to the signals at the channels $-k+1$ and/or $-k-1$ may be omitted. However, this assumption will not simplify our analysis and it would  exempt the cases where the signal strength at channels  $-k+1$ and/or $-k-1$ are strong enough to affect the energy detectors decision.

\section{Conclusions}\label{sec:Conclusions}
We studied the performance of multi-channel spectrum sensing, when the RF front-end is impaired by hardware imperfections. In particular, assuming Rayleigh fading, we provided the analytical framework for evaluating the detection and false alarm probabilities of energy detectors when LNA nonlinearities, IQI and PHN are taken into account.
Next, we extended our study to the case of a CR network, in which the SUs suffer from different levels of RF impairments, taking into consideration both scenarios of error free and imperfect reporting channels.
Our results illustrated the degrading effects of RF imperfections on the ED spectrum sensing performance, which bring significant losses in the utilization of the spectrum.  Among others, LNA non-linearities were shown to have the most detrimental effect on the spectrum sensing~performance.
Furthermore, we observed that in cooperative spectrum sensing, the sensing capabilities of the CR system are  significantly influenced by the different levels of RF imperfections of the~SUs.
Therefore, hardware constraints should be seriously taken into consideration when designing direct conversion CR RXs.

\appendix[Approximation for extended incomplete gamma function calculation]

\begin{thm}
The extended incomplete Gamma function can be approximated as
\begin{align}
\Gamma\left(a,x,b,1 \right) \approx \sum_{n=0}^{N}\frac{\left(-b\right)^{n}}{n!}\Gamma\left(a-n,x\right),\label{Eq:appr_ext_gamma}
\end{align}
with an approximation error upper-bounded by
\begin{align}
\epsilon\left(a,x,b,N\right) =  \exp\left(b\right) \Gamma\left(a-N-1,x\right)  \frac{\gamma\left(N+1,b\right)}{\Gamma\left(N+1\right)}.
\label{approximation_error}
\end{align}
\end{thm}
\begin{IEEEproof}
The extended incomplete Gamma function can be expanded in terms of the incomplete Gamma function as \cite[Eq. (6.54)]{B:chaudhry2001class}
\begin{align}
\Gamma\left(a,x,b,1 \right) = \sum_{n=0}^{\infty}\frac{\left(-b\right)^{n}}{n!}\Gamma\left(a-n,x\right).
\end{align}
By denoting
$f\left(a,x,b,n\right) = \frac{b^{n}}{n!}\Gamma\left(a-n,x\right),$
the extended incomplete gamma function can be rewritten as
$\Gamma\left(a,x,b,1 \right) = \sum_{n=0}^{\infty}\left(-1\right)^{n} f\left(a,x,b,n\right)$.
Moreover, according to \cite[Eq. (3.84)]{B:chaudhry2001class}, the auxiliary function $f\left(a,x,b,n\right)$ is equivalent to
$f\left(a,x,b,n\right) = \frac{b^{n}}{n!}\frac{E_{n-a+1}\left(x\right)}{x^{n-a}}$,
where $E_{n}\left(x\right)$ is the exponential integral function defined in \cite[Eq. (5.1.4)]{B:Abramowitz}. Taking into consideration the property \cite[Eq. (5.1.17)]{B:Abramowitz}, it follows that for given parameters $a$, $x>0$ and $n$,
\begin{align}
\Gamma\left(a-n,x\right) \geq \Gamma\left(a-n-1,x\right),\label{eq:gamma_mono}
\end{align}
and, hence, for a given $b>0$,
\begin{align}
\lim_{n \rightarrow \infty}f\left(a,x,b,n\right)=0. \label{eq:limit}
\end{align}
Thus, the extended incomplete gamma function can be approximated by (\ref{Eq:appr_ext_gamma}) where the approximation error is given by
\begin{align}
e(a,x,b,N)=\sum_{n=N+1}^{\infty}\left(-1\right)^{n}f\left(a,x,b,n\right),
\end{align}
which can be upper-bounded, according to (\ref{eq:gamma_mono}) and (\ref{eq:limit}), as
\begin{align}
e(a,x,b,N)\leq\sum_{n=N+1}^{\infty}f\left(a,x,b,n\right)\leq \Gamma\left(a-N-1,x\right)\sum_{n=N+1}^{\infty}\frac{b^n}{n!} .
\end{align}
Hence, using \cite[Eq. (1.211/1)]{B:Gra_Ryz_Book} and \cite[Eq. (8.352/2)]{B:Gra_Ryz_Book}, the upper bound on the approximation error given by (\ref{approximation_error}) is derived.
\end{IEEEproof}

\bibliographystyle{IEEEtran}
\bibliography{IEEEabrv,References}

\end{document}